\documentclass[reprint, aip, amsmath, amssymb]{revtex4-2}

\usepackage[T1]{fontenc} 

\usepackage{amsmath,color}
\usepackage{amssymb}
\usepackage{amsfonts}
\usepackage{amsthm}
\usepackage{mathtools}

\usepackage[title]{appendix}

\usepackage{algorithm}
\usepackage{algpseudocode}
\usepackage{dsfont}

\usepackage{bbold}
\usepackage{chemformula}
\usepackage{siunitx}

\DeclarePairedDelimiterX\braket[2]{\langle}{\rangle}{#1 \delimsize\vert #2}
\DeclarePairedDelimiterX\braket3[3]{\langle}{\rangle}{#1 \delimsize\vert #2 \delimsize\vert #3}

\usepackage{accents}
\newcommand{\dbtilde}[1]{\accentset{\approx}{#1}}

\newcommand{\vE}{\mathbf{E}}

\newcommand{\ve}{\mathbf{e}}

\newcommand{\vmu}{\boldsymbol{\mu}}

\newcommand{\vxi}{\boldsymbol{\xi}}

\newcommand{\hH}{\hat{H}}

\newcommand{\avg}[1]{\left\langle #1\right\rangle}
\newcommand{\red}[1]{{\color{black} #1}}

\begin{document}
	
	\title{Cavity molecular dynamics simulations of vibrational polariton enhanced molecular nonlinear absorption}
	
	\author{Tao E. Li}%
	\email{taoli@sas.upenn.edu}
	\affiliation{Department of Chemistry, University of Pennsylvania, Philadelphia, Pennsylvania 19104, USA}

	\author{Abraham Nitzan} 
	\email{anitzan@sas.upenn.edu}
	\affiliation{Department of Chemistry, University of Pennsylvania, Philadelphia, Pennsylvania 19104, USA}
	\affiliation{School of Chemistry, Tel Aviv University, Tel Aviv 69978, Israel}

	\author{Joseph E. Subotnik}
	\email{subotnik@sas.upenn.edu}
	\affiliation{Department of Chemistry, University of Pennsylvania, Philadelphia, Pennsylvania 19104, USA}

	\begin{abstract}
		Recent experiments have observed that the chemical and photophysical properties of molecules  can be modified inside an optical Fabry--P\'erot \red{microcavity} under collective vibrational strong coupling (VSC) conditions, and such modification is currently not well understood by theory. In an effort to understand the origin of such cavity induced phenomena, some recent studies have focused on the effect of the cavity environment on the nonlinear optical response of the molecular subsystem. Here, we use a recently proposed protocol for classical cavity molecular dynamics (CavMD) simulations to  numerically investigate the linear and nonlinear response of liquid carbon dioxide under such VSC conditions following an optical pulse excitation. We find that applying a strong pulse of excitation to the lower hybrid light-matter state, i.e., the lower polariton (LP), can lead to an overall molecular nonlinear absorption which is enhanced by up to two orders of magnitude relative to the excitation outside the cavity. This polariton-enhanced multiphoton absorption also causes an ultrashort LP lifetime (0.2 ps) under strong illumination. Unlike usual polariton relaxation processes --- whereby polaritonic energy transfers directly to the manifold of singly excited vibrational dark states ---  under the present mechanism, the LP transfers energy directly to the manifold of higher vibrationally excited dark states;  these highly excited dark states subsequently relax to the manifold of singly excited states with a lifetime of tens of ps.
		Because the present mechanism is generic in nature, we expect these numerical predictions to be experimentally observed in different molecular systems and in cavities with different volumes.
	\end{abstract}

	\maketitle
	
	\section{Introduction}\label{sec:intro}
	
	Molecular polaritons, the hybrid quasi-particles stemming from strong light-matter interactions, open a new avenue to control molecular properties \cite{Herrera2019}. Particular attention has recently focused on the vibrational strong coupling (VSC) regime, where the interaction between a molecular ensemble and a cavity mode of a micron-size Fabry--P\'erot cavity results in an observed Rabi splitting of order $\sim$ 100 cm$^{-1}$ \cite{Shalabney2015,George2015,George2016}. 
	Vibrational polaritons have  been implicated  in the observed  modification of various molecular properties in the electronic ground state, including (i) changes in ground-state chemical reaction rates  \cite{Thomas2016,Vergauwe2019},  reaction pathways selectivities \cite{Thomas2019_science}, and even chemical equilibria \cite{Pang2020} without external  pumping; (ii) modification of optical nonlinearities \cite{Xiang2019Nonlinear,F.Ribeiro2018} and (iii) enhanced  intermolecular vibrational energy transfer (VET) rates \cite{Xiang2020Science}. Significantly, these modifications appear to be collective phenomena, originating from the interaction of cavity mode(s) with a large number of molecules.

	From a theoretical perspective, collective optical response  is easily understandable. In particular, vibrational Rabi splitting is easily modeled by mapping molecular vibrations and cavity modes onto  harmonic oscillators \cite{Rudin1999}. By contrast, cavity induced chemical phenomena must involve a nuanced balance of collective and individual effects, and many questions remain as to the exact origin of the observed cavity-induced chemical effects \cite{Galego2019,Li2020Origin,Campos-Gonzalez-Angulo2020,Zhdanov2020,Scholes2020}. 
	Indeed, so named "chemical catalysis" under VSC cannot be explained by a simple transition-state theory of chemical reactions because the potential of mean force along a reaction pathway is unchanged inside the cavity  \cite{Li2020Origin,Campos-Gonzalez-Angulo2020,Zhdanov2020}. Currently, descriptions  of polariton  relaxation \cite{Xiang2018, Xiang2019State, Grafton2020},  polariton-enhanced optical nonlinearity \cite{Ribeiro2020} and intermolecular VET rates \cite{Xiang2020Science}  under VSC are largely limited to either phenomenological master equations (which rely on the parameter fitting from experiments) or single-molecule analytical models that cannot address the collective aspects of the observed phenomena.
	
	For a different perspective, we have recently proposed a numerical scheme to study cavity effects --- cavity molecular dynamics simulations (CavMD)  \cite{Li2020Water}, in which classical dynamics is used to propagate coupled photon-nuclei dynamics for realistic molecular models. Compared with  recently proposed theoretical methods in polariton chemistry which mainly focus on  electronic structure calculations \cite{Herrera2016,Flick2017,Luk2017,Groenhof2019,Schafer2020,Mandal2020} for a molecular electronic transition strongly coupled to a cavity mode (i.e., under electronic strong coupling), our approach serves as an affordable classical tool for describing  VSC for a large ensemble of realistic molecules with full atomic resolution. Our approach  \cite{Li2020Water} captures the asymmetric Rabi splitting   \cite{Vergauwe2019} under VSC and even vibrational ultrastrong coupling when the \ch{O-H} stretch mode of liquid water is strongly coupled to a cavity mode.

	An important result from CavMD simulations of liquid water is that the static equilibrium properties of \ch{H2O} are completely unchanged inside versus outside the cavity, and
	dynamic response functions (evaluated in the linear response regime) of individual molecules also show very little or no effect. Thus, one must conclude that  any significant VSC modification, e.g., VSC "catalysis" and (or) the acceleration of an intermolecular VET rate, cannot be explained from an equilibrium or linear-response point of view. This conclusion suggests that cavity effects that apply to individual molecules may reflect cavity modification of the nonequilibrium dynamics of the molecular subsystem.

	In this paper we will present the results of such nonequilibrium CavMD simulations, looking at how a molecular system behaves following the application of an external pulse that excited a vibrational polariton; our goal is to explore if and how cavity effects may modify molecular nonequilibrium properties. We will show that nonequilibrium CavMD simulations can recover  experimental observations such as polariton relaxation to vibrational dark modes \cite{Xiang2018} and a delay in the population gain of the singly excited states of vibrational dark modes  after pumping  the lower polariton (LP) \cite{Xiang2019State}; moreover, such simulations also predict  an intriguing process whereby a LP can enhance the molecular nonlinear absorption of light by up to two orders of magnitude, leading to very  large molecular populations of highly excited vibrational states.

\begin{figure}
		\centering
		\includegraphics[width=1.0\linewidth]{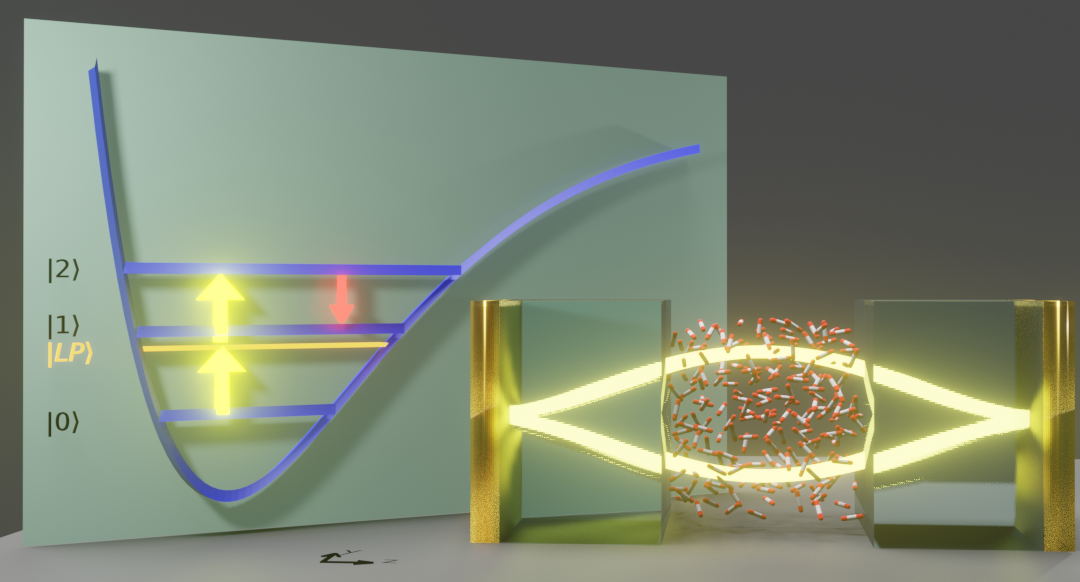}
		\caption{Sketch of the simulation setup where a large collection of \ch{CO2} molecules is confined between a pair of metallic mirrors; see Sec. \ref{sec:simu} for details. As shown in the left cartoon, the main finding of this manuscript is that strongly exciting the LP in a cavity can greatly enhance the multiphoton nonlinear absorption of molecules relative to that outside the cavity. This enhancement is largest when twice the LP frequency approximately matches the vibrational $0\rightarrow 2$ transition. After one directly excites the LP, and indirectly excites a localized vibrational state with $v$ = 2, there is subsequently a gradual transfer of population to the first vibrationally excited state with a timescale that is much slower than the LP lifetime.}
		\label{fig:toc}
	\end{figure}

	The model system studied is an ensemble of carbon dioxide molecules, where the \ch{C=O} asymmetric stretch mode of liquid \ch{CO2} is nearly resonant with a cavity mode and forms lower and upper polaritons (UP) under VSC; see Fig. \ref{fig:toc} for the simulation setup. This  model system resembles experimental systems studied recently  with weak intermolecular interactions (such as \ch{W(CO)6}) \cite{Xiang2018,Xiang2019State,Xiang2019}.
	
	\section{Method}\label{sec:method}
	Here, we outline the theoretical considerations that underlie our  CavMD simulations. A detailed account is given in Ref \cite{Li2020Water}. Under the Born--Oppenheimer approximation,  the full-quantum photon-electron-nuclei Hamiltonian is projected onto the electronic ground state, which leads to a quantum Hamiltonian for the coupled photon-nuclei system only:
	\begin{subequations}\label{eq:H}
		\begin{equation}
		\begin{aligned}
		\hH_\text{QED}^{\text{G}} = \ & \hH_{\text{M}}^{\text{G}} 
		+ \hH_{\text{F}}^{\text{G}} .
		\end{aligned}
		\end{equation}
		Here, $\hH_{\text{M}}^{\text{G}}$ denotes the conventional ground-state molecular Hamiltonian:
		\begin{align}\label{eq:H_QED_G-2}
		\hH_{\text{M}}^{\text{G}} = \sum_{n=1}^{N}\left(\sum_{j\in n} \frac{\hat{\mathbf{P}}_{nj}^2}{2 M_{nj}} + \hat{V}^{(n)}_{g}(\{\hat{\mathbf{R}}_{nj}\})\right) + 
		\sum_{n=1}^{N}\sum_{l>n}
		\hat{V}_{\text{inter}}^{(nl)}
		\end{align}
		where  $\hat{\mathbf{P}}_{nj}$, $\hat{\mathbf{R}}_{nj}$, and $M_{nj}$  denote the momentum operator, position operator, and mass for the $j$-th nucleus in molecule $n$, $\hat{V}^{(n)}_{g}$ denotes the intramolecular potential for molecule $n$, and $\hat{V}_{\text{inter}}^{(nl)}$ denotes the intermolecular interactions between molecule $n$ and $l$.
		$\hH_{\text{F}}^{\text{G}}$ denotes the field-related Hamiltonian \cite{Cohen-Tannoudji1997, Li2020Origin, Haugland2020}:
		\begin{equation}\label{eq:H_QM}
		\begin{aligned}
		\hH_{\text{F}}^{\text{G}} & = \sum_{k,\lambda}
		\frac{\hat{\widetilde{p}}_{k, \lambda}^2}{2 m_{k, \lambda}}  + 
		\frac{1}{2} m_{k,\lambda}\omega_{k,\lambda}^2 \Bigg (
		\hat{\widetilde{q}}_{k,\lambda}  + \sum_{n=1}^{N} \frac{\hat{d}_{ng, \lambda}}{\omega_{k,\lambda}\sqrt{\Omega\epsilon_0 m_{k,\lambda}}} 
		\Bigg )^2 
		\end{aligned}
		\end{equation}
		where $\hat{\widetilde{p}}_{k, \lambda}$, $\hat{\widetilde{q}}_{k, \lambda}$, $\omega_{k,\lambda}$, and $m_{k, \lambda}$ denote the momentum operator, position operator, frequency, and the auxiliary mass for each cavity photon mode with wave vector $\mathbf{k}$ and polarization direction $\vxi_\lambda$. Note that the use of the auxiliary mass does not alter any dynamics and is necessary only because most MD packages require such mass; since the full light-matter coupling term in Eq. \eqref{eq:H_QM} is proportional to $\hat{q}_{k,\lambda} \equiv \sqrt{m_{k,\lambda}}\hat{\widetilde{q}}_{k,\lambda}$, where $\hat{q}_{k,\lambda}$ is the standard mass-reduced photonic position operator, and the final dynamics of the physical $\hat{q}_{k,\lambda}$ operator and any molecular operator are not influenced by the magnitude of the auxiliary mass.
		\footnote{As a practical matter, even though the raw value of $\hat{\widetilde{q}}_{k,\lambda}$ is different from the raw value of $\hat{q}_{k,\lambda}$,  the spectrum and the energy of the photon can be calculated with either $\hat{q}_{k,\lambda}$  or $\hat{\widetilde{q}}_{k,\lambda}$ and the same results can be obtained.}
		$\Omega$ denotes the volume for the microcavity, $\epsilon_0$ denotes the vacuum permittivity, and $\hat{d}_{ng, \lambda}$ denotes the electronic ground-state dipole operator for molecule $n$ projected along the direction of $\vxi_\lambda$.
		In Eq. \eqref{eq:H_QM}, we can define
		\begin{equation}\label{eq:H_QM_coupling}
			\varepsilon_{k,\lambda}\equiv \sqrt{m_{k,\lambda}\omega_{k,\lambda}^2/\Omega\epsilon_0}
		\end{equation}	
		to characterize the coupling strength between each cavity photon and individual molecules.
		Note that Eq. \eqref{eq:H_QM} assumes the long-wave approximation, i.e., the molecular ensemble is assumed to be much smaller than the wavelength of the cavity mode. 
		Eq. \eqref{eq:H_QM} is exact when the cavity volume is large, e.g.,  in a microcavity where collective VSC is studied, otherwise an additional self-dipole fluctuation term will also emerge due  to the quantum nature of electrons \cite{Li2020Origin}.
		Compared with most model Hamiltonians such as the Tavis--Cummings Hamiltonian, the most different feature in Eq. \eqref{eq:H_QM} is the inclusion of the self-dipole term, i.e., the term that is quadratic in $\hat{d}_{ng,\lambda}$. We emphasize that including this self-dipole term is critical in CavMD simulations because this term preserves gauge invariance, contributes to the asymmetry of Rabi splitting \cite{Li2020Water}, and most importantly, influences the long-time molecular dynamics and the reliability of CavMD results. For example, for thermal equilibrium simulations, including the self-dipole term guarantees that  static molecular properties will be unchanged under VSC \cite{Li2020Water}, while neglecting such a  term will cause  numeric artifacts (such as changed static molecular properties). See also Ref. \red{\cite{Schafer2020,Feist2020}} which discusses the importance of the self-dipole term in other contexts.
	\end{subequations}
	
	After reducing all of the operators in Eq. \eqref{eq:H} to classical observables and also applying periodic boundary conditions for the molecules, we arrive at  equations of motion for the coupled photon-nuclei system:
	\begin{subequations}\label{eq:EOM_MD_PBC}
		\begin{align}
		M_{nj}\ddot{\mathbf{R}}_{nj} &= \mathbf{F}_{nj}^{(0)}  + \mathbf{F}_{nj}^{\text{cav}} + \mathbf{F}_{nj}^{\text{ext}}(t)
		\\
		m_{k,\lambda}\ddot{\dbtilde{q}}_{k,\lambda} &= - m_{k,\lambda}\omega_{k,\lambda}^2 \dbtilde{q}_{k,\lambda}
		-\widetilde{\varepsilon}_{k,\lambda} \sum_{n=1}^{N_{\text{sub}}}d_{ng,\lambda}
		\end{align}
	Here, the subscript $nj$ denotes the $j$-th atom in molecule $n$;  $\mathbf{F}_{nj}^{(0)} = - \partial V_g^{(n)} / \partial \mathbf{R}_{nj} - \sum_{l\neq n} \partial V_{\text{inter}}^{(nl)}/\partial \mathbf{R}_{nj}$ denotes the molecular part of the force on each nuclei; 
	$\mathbf{F}_{nj}^{\text{cav}} = - 
	\sum_{k,\lambda}
	\Large(
	\widetilde{\varepsilon}_{k,\lambda} \dbtilde{q}_{k,\lambda}
	+ \allowbreak \frac{\widetilde{\varepsilon}_{k,\lambda}^2}{m_{k,\lambda} \omega_{k,\lambda}^2} \sum_{l=1}^{N_{\text{sub}}} d_{lg,\lambda}
	\Large) 
	\frac{\partial d_{ng, \lambda}}{\partial \mathbf{R}_{nj}}$ denotes the cavity force on each nuclei;
	$\mathbf{F}_{nj}^{\text{ext}}(t) = -Q_{nj} \vE_{\text{ext}}(t)$ denotes an external driving force acting on each nuclei with partial charge $Q_{nj}$ under the pumping of a time-dependent electric field $\vE_{\text{ext}}(t)$. Note that this $\mathbf{F}_{nj}^{\text{ext}}(t)$ was not introduced in Ref. \cite{Li2020Water} and is included here to represent the optical pulse excitation that leads to the molecular nonequilibrium response.
	\footnote{Although in principle an external pulse can also directly excite cavity photons, this feature is not included in Eq. \eqref{eq:EOM_MD_PBC}. Including such an excitation would  not qualitatively change the simulation results in this manuscript since pumping either the molecular or the photonic part  equivalently pumps polaritons. More importantly, in order to include this feature, one would need to introduce a new phenomenological term, namely the effective transition dipole of cavity photons, the magnitude of which varies for different cavities. Therefore, introducing this term would bring an additional  manipulatable parameter and would hinder the universality of our simulation results: in Sec. \ref{sec:result_PBC} we will show that the presented results are universal for cavities with different volumes.}
	In order to apply periodic boundary conditions, in Eq. \eqref{eq:EOM_MD_PBC}, we have redefined $\dbtilde{q}_{k,\lambda} =  \tilde{q}_{k,\lambda} / \sqrt{N_{\text{cell}}}$, where $N_{\text{cell}}$ denotes the number of the periodic simulation cells for molecules, and also an effective coupling strength
	\begin{equation}
		\widetilde{\varepsilon}_{k,\lambda} = \sqrt{N_{\text{cell}}} \varepsilon_{k,\lambda},
	\end{equation}
	 where $\varepsilon_{k,\lambda}$, the true coupling strength between each cavity photon and individual molecules, has been defined in Eq. \eqref{eq:H_QM_coupling}.
	 By invoking periodic boundary conditions as above, CavMD simulations can yield the same Rabi splitting  when calculating the consequence of photons interacting with molecules in a simulation cell as found in the original system.
	 With the number of molecules in a single simulation cell denoted as  $N_{\text{sub}}$, the total number of molecules is $N=N_{\text{sub}}N_{\text{cell}}$. Note that when CavMD is used to reproduce experimental observations such as polariton relaxation, given the value of $N_{\text{sub}}$, $\widetilde{\varepsilon}_{k,\lambda} \propto \sqrt{N_{\text{cell}}}$ 
	can be chosen by fitting the experimentally observed Rabi splitting $\Omega_N$; since  $\Omega_N \propto \sqrt{N}$, $\widetilde{\varepsilon}_{k,\lambda} \propto \sqrt{N_{\text{cell}}} \propto \Omega_N / \sqrt{N_{\text{sub}}}$. 
	\end{subequations}
	Also note that if CavMD is used to simulate VSC phenomena in Fabry--P\'erot microcavities with $N$ molecules using parameters designed to \red{recapitulate} an experimental Rabi splitting $\Omega_N$, it is necessary to check the dependence on periodic boundary conditions (or the choice of $N_{\text{sub}}$) to make sure that any observed dynamics are not an artifact of the simulation; after all, the effective coupling strength for molecules  $\widetilde{\varepsilon}_{k,\lambda}$ has been amplified by a factor of $\sqrt{N_{\text{cell}}}$ relative to the true coupling strength $\varepsilon_{k,\lambda}$. 
	In Sec. \ref{sec:result_PBC}, we will study how our CavMD results depend on $N_{\text{sub}}$, all while keeping the Rabi splitting $\Omega_N$ fixed and adjusting $\widetilde{\varepsilon}_{k,\lambda}$  accordingly.  For such calculations,  the asymptotic results when $N_{\text{sub}}$ approaches a macroscopic number should correspond to  Fabry--P\'erot microcavities. For the same calculations,
	it may also be helpful to imagine a physical experiment with $N_{\text{cell}}=1$ (so that $\widetilde{\varepsilon}_{k,\lambda} = \varepsilon_{k,\lambda}$ becomes the true coupling strength); here, one can interpret our CavMD dynamics as reliably reporting on cavities with different effective volumes (and therefore different effective molecular numbers $N$).
	
	\subsection{Molecular spectroscopy}
	Below we will calculate two different spectroscopic response functions: the global infrared (IR) absorption spectrum and its "local" correspondence which is the spectrum obtained if the molecules respond to light individually. The former is calculated by Fourier transforming the dipole auto-correlation function \cite{McQuarrie1976,Gaigeot2003,Habershon2008,Nitzan2006}:
	\begin{equation}\label{eq:IR_equation_cavity}
	\begin{aligned}
	n(\omega)\alpha(\omega) &= \frac{\pi \beta \omega^2}{2\epsilon_0 V c} \frac{1}{2\pi}   \int_{-\infty}^{+\infty} dt \ e^{-i\omega t}  \\
	& \times \avg{\sum_{i=x, y}\left(\vmu_S(0)\cdot \ve_i\right) \left(\vmu_S(t)\cdot \ve_i\right)} 
	\end{aligned}
	\end{equation}
	while the latter is defined by \begin{equation}\label{eq:localIR_equation_cavity}
		\begin{aligned}
			n(\omega)\alpha_{\text{local}}(\omega) &= \frac{\pi \beta \omega^2}{2\epsilon_0 V c} \frac{1}{2\pi}   \int_{-\infty}^{+\infty} dt \ e^{-i\omega t} \\
			&\times  \avg{\frac{1}{N_{\text{sub}}}\sum_{n=1}^{N_{\text{sub}}} \vmu_n(0)\cdot \vmu_n(t)} 
		\end{aligned}
	\end{equation}
	In Eqs. \eqref{eq:IR_equation_cavity} and \eqref{eq:localIR_equation_cavity}, $\alpha(\omega)$ or $\alpha_{\text{local}}(\omega)$  denotes the absorption coefficient, $n(\omega)$ denotes  the refractive index, $V$ is the volume of the system (i.e., the simulation cell),
	$\ve_i$ denotes the unit vector along direction $i=x, y$, and $\vmu_S(t)$ denotes the \textit{total} dipole moment of the molecules at time $t$, where $\vmu_S(t) = \sum_{nj} Q_{nj} \mathbf{R}_{nj}(t)$. Note that in our force field calculations, the partial charges ($Q_{nj}$, the true nuclear charge for nucleus $nj$ plus the electronic shielding effect) of nuclei are fixed; see Appendix \ref{App:force_field} for the values of partial charges. $\vmu_n(t) = \sum_{j} Q_{nj} \mathbf{R}_{nj}(t)$
	denotes the dipole moment for molecule $n$.  In Eq. \eqref{eq:IR_equation_cavity}, the summation over $x$ and $y$ is a summation over the two possible polarizations of the relevant cavity mode of the $z$-oriented cavity (see Fig. \ref{fig:toc} for the simulation setup). In Eq. \eqref{eq:localIR_equation_cavity}, the inner product implies that a summation over three dimensions has been applied, though summing over only two dimensions $x,y$ yields  the same local IR spectra up to a prefactor.

	It is important to note that $\alpha(\omega)$ of Eq. \eqref{eq:IR_equation_cavity} represents the molecular IR absorption (for a molecular sample much smaller than the wavelength of light). Its local correspondence, $\alpha_{\text{local}}(\omega)$ of Eq. \eqref{eq:localIR_equation_cavity}, corresponds to the absorption of a fictitious molecular system in which each molecule responds to the field individually. Thus, Eq. \eqref{eq:IR_equation_cavity} describes the collective behavior of the molecular dipole system, while Eq. \eqref{eq:localIR_equation_cavity} provides information about the dynamics of individual molecules. 
	Just as the collective bright and dark modes of the molecular ensemble can be expressed as a linear combination of individual molecular modes, the individual response can also be expressed as a linear combination of the collective modes --- note, though, that the latter is  predominately dominated by the dark modes. For instance, if $N_{\text{sub}} = 216$ is taken during the simulation, the bright state contribution to the local IR spectrum in Eq. \eqref{eq:localIR_equation_cavity} is negligibly small, equal to  $1/N_{\text{sub}} = 1/216$. Therefore, one can roughly interpret the local IR spectrum as reporting  dark-mode dynamics.
	
	Lastly, a few words are appropriate as far as interpreting the response functions above. Quantum-mechanically, one calculates absorption and emission spectra differently; even though one must propagate the same Hamiltonian in both cases, absorption and emission can be differentiated by their respective initial conditions and the fact that quantum correlation functions are not time-reversible, $\avg{\hat{A}\hat{B}(t)} \neq \avg{\hat{A}\hat{B}(-t)}$.  Classically, correlation functions (and their spectra) report on the energy present in a given mode, which cannot be naturally dissected into absorption or emission components, and this inability can lead to some confusion when analyzing MD simulations and looking to connect with experimental spectra (that reflect true quantum mechanical dynamics).  Nevertheless, by following vibrational energy evolves in time during a given trajectory, we will be able to semiclassically rationalize how the cavity mode, the bright mode, and the dark modes relax and exchange energy.

	\section{Simulation details}\label{sec:simu}
	
	A sketch of the cavity structure used in our simulation is plotted in Fig. \ref{fig:toc}. 
	The cavity is placed along the $z$-axis. In addition to the (assumed perfect) mirrors that define the Fabry--P\'erot cavity, we assume that additional parallel inert layers (e.g., \ch{SiO2}, which has been used in references \cite{Thomas2016}) confine the molecular system in a region small enough to   guarantee the validity of the long wave approximation and also far enough from the mirrors so that image charge effects can be disregarded. Both assumptions are made in writing the model Hamiltonian \eqref{eq:H}.
	Only one cavity mode with frequency near the \ch{C=O} asymmetric stretch is considered. This cavity mode contains two polarization directions along the $x$ and $y$ directions. We set the auxiliary mass for the cavity mode as $m_{k, \lambda} = 1$ a.u. (atomic units) though, as mentioned above, this mass does  not affect any  dynamics. Under periodic boundary conditions, we simulate 216 \ch{CO2} molecules in a cubic cell with cell length 24.292 \si{\angstrom} (45.905 a.u.); the density of the liquid \ch{CO2} is 1.101 $\text{g/cm}^{3}$.  Unless stated otherwise, by default, we set the cavity mode frequency as 2320 cm$^{-1}$ with an effective coupling strength $\widetilde{\varepsilon} = 2\times 10^{-4}$ a.u.. Compared with VSC experiments for which usually $N  = 10^9 \sim 10^{11}$ molecules are involved, our choice of the effective coupling strength $\widetilde{\varepsilon} \propto \sqrt{N_{\text{cell}}}$ should correspond to the involvement of $N_{\text{cell}} =  10^7 \sim 10^9$  periodic simulation cells. 
	
	When calculating equilibrium properties, we perform simulations as follows.
	At 300 K, we first run the simulation for 150 ps to guarantee thermal equilibrium under a NVT (constant particle number, volume, and temperature) ensemble where a Langevin thermostat with a lifetime (i.e., inverse friction) of 100 fs is applied to the momenta of all particles (nuclei + photons). The resulting equilibrium configurations are used as starting points for  40 consecutive NVE (constant particle number, volume, and energy) trajectories of length 20 ps. At the beginning of each trajectory the velocities are resampled by a Maxwell-Boltzmann distribution under 300 K. The intermolecular Coulombic interactions are calculated by an Ewald summation.  The simulation step is set as $0.5$ fs and we store the snapshots of trajectories every 2 fs. 
	
	When performing nonequilibrium simulations under an external pulse, we start each simulation with an equilibrium geometry, which is chosen from the starting configurations of the above 40 NVE trajectories. Each nonequilibrium trajectory is run for 100 ps under a NVE ensemble and  the  physical properties are calculated by averaging over these 40 nonequilibrium trajectories. Note that the use of NVE trajectories when calculating equilibrium or nonequilibrium physical properties implies the assumption that cavity losses are small on the timescale of simulations and can be ignored. This situation is usually valid when considering Fabry--P\'erot microcavities where the cavity loss lifetime usually takes $\sim 5$ ps, while polariton relaxation to vibrational dark modes usually occurs on a timescale $< 5$ ps; see Sec. \ref{sec:result_lifetime} Fig. \ref{fig:lifetime_all} and also experiments \cite{Xiang2019State}. Of course, for cavities with larger losses or when the polariton relaxation process occurs on a longer timescale, the polariton relaxation might be accelerated and 
	the resulting linear and nonlinear vibrational dark-mode energy absorption (in Fig. \ref{fig:dynamics_CO2_LP}) would then also be weakened.
	
	The form of the external pulse is taken as follows:
	\begin{equation}\label{eq:Ex}
	\vE_{\text{ext}}(t) = E_0 \cos\left(\omega t + \phi\right) \ve_x
	\end{equation}
	where the phase $\phi \in [0, 2\pi)$ is set as random. This pulse is turned on at $t_{\text{start}} = 0.1$ ps and is turned off at $t_{\text{end}} = 0.6$ ps.
	Below we will show results obtained after weak pumping, i.e., $E_0 =  3.084 \times 10^{6}$ V/m ($6\times 10^{-4}$ a.u.) and the input pulse fluence $F = \frac{1}{2}\epsilon_0 c E_0^2 (t_{\text{end}} - t_{\text{start}}) = 6.32$ mJ/cm$^2$, and also after strong pumping, i.e., $E_0 =  3.084 \times 10^{7}$ V/m ($6\times 10^{-3}$ a.u.) and $F = 632$ mJ/cm$^2$.
	The choice of an $x$-polarized pumping pulse implies that molecules with dipole component in the $x$ direction can be excited, and because the cavity sits along the $z$-direction, the $x$-polarized  pulse (in Eq. \eqref{eq:Ex}) can excite the polariton associated with the cavity mode polarized along the same $x$-direction.
	Throughout this manuscript, we will refer to exciting the polaritons as exciting the molecular ensemble with such a $x$-polarized pulse.  Similarly,  polaritons can also be excited with  a $y$-polarized pulse.
	
	\begin{figure*}
		\centering
		\includegraphics[width=1.0\linewidth]{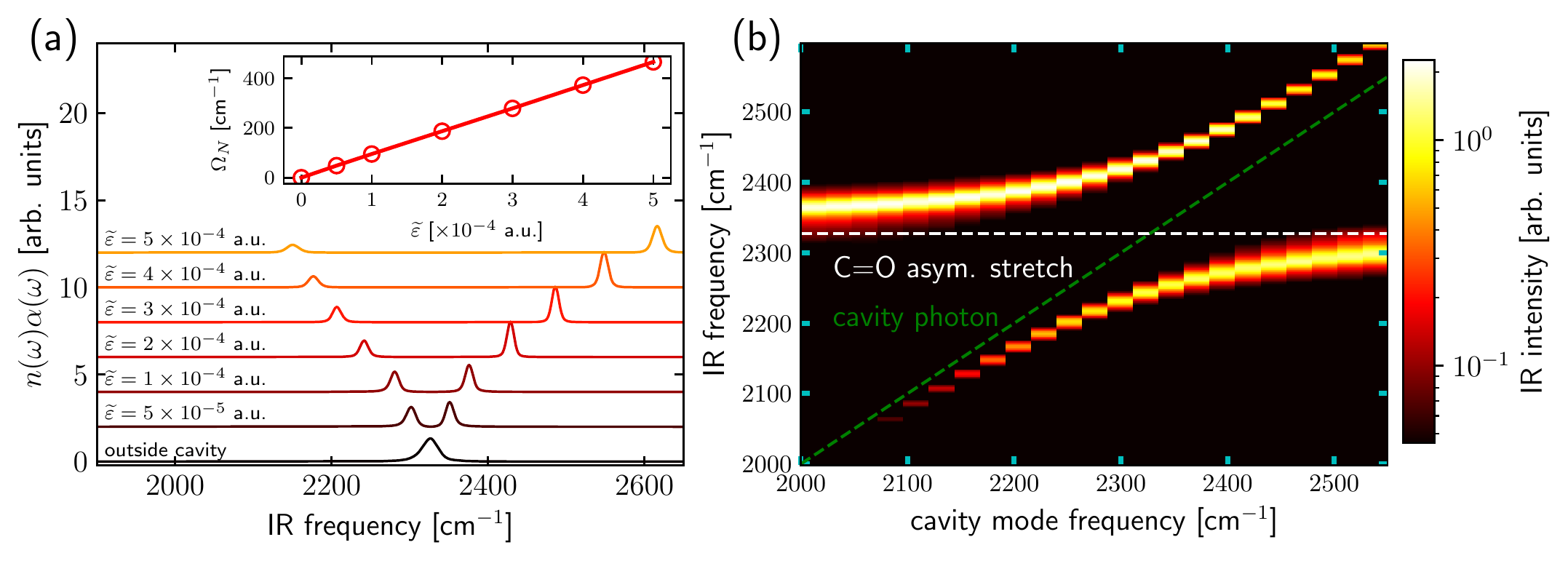}
		\caption{Rabi splitting from equilibrium simulations. (a) IR spectrum as a function of the increased (from bottom to top) effective coupling strength $\widetilde{\varepsilon}$  when the cavity mode frequency (2320 cm$^{-1}$) is nearly resonant with the \ch{C=O} asymmetric stretch (2327 cm$^{-1}$). Inset: The linear relationship between Rabi splitting ($\Omega_\text{N}$) and $\widetilde{\varepsilon} \propto \sqrt{N}$. (b) Avoided crossing of the IR spectrum as a function of the cavity mode frequency given $\widetilde{\varepsilon} = 2\times 10^{-4}$ a.u.. The dashed white line denotes the asymmetric \ch{C=O} stretch outside a cavity, the dashed green line denotes the cavity mode, and the intensities of IR spectra are plotted on a logarithmic scale; see the colorbar. Note that the IR spectra here are calculated from equilibrium trajectories by evaluating the dipole auto-correlation function in Eq. \eqref{eq:IR_equation_cavity}; see Sec. \ref{sec:simu} for other simulation details.}
		\label{fig:IR_CO2}
	\end{figure*}
	
	For the force field of \ch{CO2} (see Appendix \ref{App:force_field} for details),
	we largely follow Ref. \cite{Cygan2012}. The only difference is that while Ref. \cite{Cygan2012} uses a harmonic potential for the \ch{C=O} bond, we change this harmonic potential to the following anharmonic form:
	\begin{equation}\label{eq:VCO}
	V_{\ch{CO}}(r) = D_r \left[\alpha_r^2 \Delta r^2
	- \alpha_r^3 \Delta r^3
	+ \frac{7}{12}\alpha_r^4 \Delta r^4
	\right]
	\end{equation}
	where $\Delta r = r - r_{\text{eq}}$. Eq. \eqref{eq:VCO} is a fourth-order Taylor expansion of a Morse potential $V_{\text{M}}(r) = D_r[1 - \exp(-\alpha_r\Delta r)]^2$.
	The parameters are taken as follows: $r_{\text{eq}} =$ 1.162 \si{\angstrom} (2.196 a.u.), $D_r = $ 127.13 kcal mol$^{-1}$ (0.2026 a.u.), and $\alpha_r = $ 2.819  \si{\angstrom}$^{-1}$ (1.492 a.u.) are chosen to fit the harmonic potential used in Ref. \cite{Cygan2012} in the harmonic limit and the value of $D_r$ takes the bond dissociation energy of \ch{O=CO} at room temperature \cite{darwent1970bond}.
	A comparison of Eq. \eqref{eq:VCO}, the Morse potential, and the harmonic limit is plotted in Appendix Fig. \ref{fig:anharmonic_potential}. When the \ch{C=O} bond energy is smaller than $\sim$ 0.05 a.u. ($10^4$ cm$^{-1}$), Eq. \eqref{eq:VCO} agrees with the Morse potential very well. However, the Morse potential is not used for the present simulation as we wish to avoid any potential molecular dissociation. 
	Note that the use of an anharmonic  potential (rather than a harmonic one) is critical for the present paper because, as will be shown in Sec. \ref{sec:result_LP}, polariton-enhanced molecular nonlinear absorption will rely on a non-uniform distribution of molecular energy level spacing.
	This reliance also highlights the fact that, in order to capture (at least some) nontrivial VSC phenomena, it is crucial to use realistic molecules instead of conventional harmonic models.

	As far as the technical details are concerned, the initial configuration is prepared with PACKMOL \cite{Martinez2009}, the CavMD scheme is implemented by modifying the I-PI package \cite{Kapil2019}, and the nuclear forces are evaluated by calling LAMMPS \cite{Plimpton1995}.
	A toolkit including source code, input and post-processing scripts, and the corresponding tutorials is available on Github \cite{TELi2020Github}.

	\section{Results and Discussion}\label{sec:results}

	\subsection{Rabi splitting and avoided crossing}

	In Fig. \ref{fig:IR_CO2}a, we plot the IR spectrum obtained from Eq. \eqref{eq:IR_equation_cavity} for different values of the effective coupling strength $\widetilde{\varepsilon}$; the value of $\widetilde{\varepsilon}$ (in a.u.) are labeled on each lineshape and the case of molecular system outside the cavity corresponds to $\widetilde{\varepsilon} = 0$. The Rabi splitting seen in Fig. \ref{fig:IR_CO2}a confirms the existence of a strong coupling between the \ch{C=O} asymmetric stretch and the cavity mode at frequency 2320 cm$^{-1}$.
	Note that with the periodic boundary conditions applied during CavMD simulations, the effective coupling strength $\widetilde{\varepsilon}$ scales as $\widetilde{\varepsilon} \propto \sqrt{N}$ with the total number of molecules $N$; see Sec. \ref{sec:method} for details. Therefore, increasing  $\widetilde{\varepsilon}$ provides a simple way to study the Rabi splitting as a function of the total number of molecules $N$. 
	As shown in Fig. \ref{fig:IR_CO2}a, unlike the lineshape outside a cavity (bottom), inside a cavity a pair of UP and LP is formed under VSC  and the Rabi splitting increases with  $\widetilde{\varepsilon}$.
	The inset plots the Rabi frequency, or the frequency difference between the UP and LP peaks, as a function of  $\widetilde{\varepsilon}\propto \sqrt{N}$. Here, a linear relationship is observed, which agrees with both analytical models of coupled harmonic oscillators and many experiments. Note that the asymmetry in the positions and amplitudes of the UP and LP seen in Fig. \ref{fig:IR_CO2}a is discussed in detail in Ref. \cite{Li2020Water}.
	
	Fig. \ref{fig:IR_CO2}b plots Rabi splitting as a function of the cavity mode frequency for the condition $\widetilde{\varepsilon} = 2\times 10^{-4}$ a.u.. When the 
	cavity mode frequency is highly negatively detuned --- i.e., when the cavity frequency is much smaller than the bare \ch{C=O} asymmetric stretch at 2327 cm$^{-1}$ [white dashed line (which is very close to the experimental value, e.g., 2333 cm$^{-1}$ in Ref. \cite{Seki2009})] --- the UP is close to the bare \ch{C=O} asymmetric stretch and its character is dominated by this molecular vibrational mode; at the same time, the LP is close to the cavity mode frequency (green dashed line) and its character is dominated by the  cavity  mode.  By contrast, when the cavity mode frequency is highly positively detuned, the LP (UP) is mostly contributed by the molecular vibration (cavity mode). The avoided crossing seen between these limits expresses the Rabi splitting  that measures the collective coupling strength.
	
	\begin{figure*}
		\centering
		\includegraphics[width=1.0\linewidth]{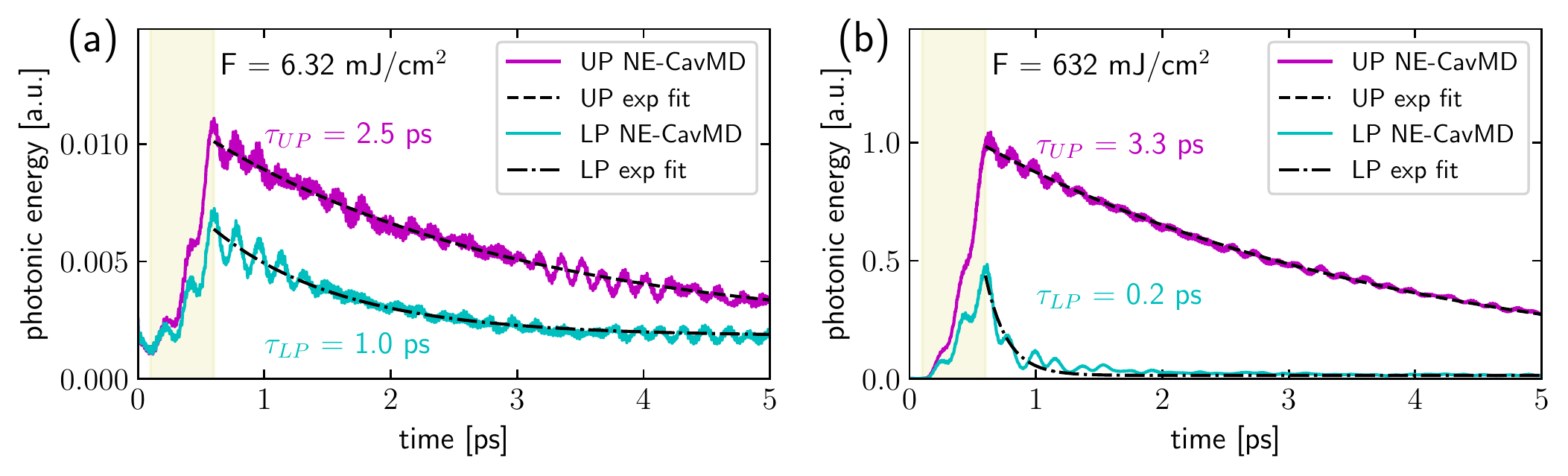}
		\caption{Time-resolved dynamics for the cavity photon energy after resonantly exciting the UP (magenta; peaked at 2428 cm$^{-1}$) or LP (cyan; peaked at 2241 cm$^{-1}$) with a (a) weak or (b) strong pulse $\vE(t) = E_0\cos(\omega t + \phi) \ve_x$ of 0.5 ps duration. The pulse fluence is $F=6.32$ mJ/cm$^{-1}$ or $F=632$ mJ/cm$^{-1}$, respectively. The  cavity mode frequency is $\omega_c = 2320 \ \text{cm}^{-1}$ and effective coupling strength is $\widetilde{\varepsilon}=2\times 10^{-4}$ a.u.; see Sec. \ref{sec:simu} for other details. The polariton lifetimes ($\tau_{\text{LP}}$ and $\tau_{\text{UP}}$ labeled at each subplot) are obtained by an exponential fit of the decaying energy of the cavity photon after the pulse.}
		\label{fig:lifetime_all}
	\end{figure*}
	
	\subsection{Polariton relaxation and ultrashort LP lifetime}\label{sec:result_lifetime}

	Next we use nonequilibrium CavMD simulations to explore the polariton relaxation following pulse excitation of the UP or the LP  (see Eq. \eqref{eq:Ex}). Because polaritons are hybrid light-matter quasi-particles, their relaxation can be captured by  monitoring either the photonic or matter side. The photonic energy ($\sum_{\lambda=x,y}m_{k,\lambda}\omega_{k,\lambda} \dbtilde{q}_{k,\lambda}^2/2 + \dbtilde{p}_{k,\lambda}^2/2m_{k,\lambda}$) is simpler to calculate and is used in this calculation \footnote{Note that consistent polaritonic relaxation dynamics can also be captured by evaluating the square norm of the molecular total dipole moment ($\avg{|\vmu_S(t)|^2} - |\avg{\vmu_S(t)}|^2$).}.
	
	Fig. \ref{fig:lifetime_all}a shows the time-resolved photonic energy, in a system where the cavity mode frequency is set to 2320 cm$^{-1}$ and effective coupling strength $\widetilde{\varepsilon}=2\times 10^{-4}$ a.u., after resonantly exciting the UP (magenta; peaked at 2428 cm$^{-1}$) or LP (cyan; peaked at 2241 cm$^{-1}$) with a weak incoming pulse $\vE(t) = E_0\cos(\omega t + \phi) \ve_x$ (with fluence $F = 6.32$ mJ/cm$^2$), where the yellow-shadowed region denotes the 0.5 ps time window during which the pulse is applied. 
	The fast oscillations (with a period $\sim 0.2$ ps) of the cavity photon signals appears to reflect  oscillations between the cavity photons and vibrational bright mode, as here the Rabi splitting is $187$ cm$^{-1} = 0.18$ ps.
	By fitting the energy decay observed after the pulse to an exponential function, the polariton lifetime can be captured: the UP lifetime is $\tau_{\text{UP}}=2.5$ ps and the LP lifetime is $\tau_{\text{LP}} = 1.0$ ps. 	
	By contrast, under a strong incoming pulse  ($F = 632$ mJ/cm$^2$), the same plot in Fig. \ref{fig:lifetime_all}b shows that while the UP lifetime is largely unchanged ($\tau_{\text{UP}} = 3.3$ ps), the LP shows an ultrashort decay with lifetime greatly reduced to $\tau_{\text{LP}} = 0.2$ ps. 
	Note that this residual decay follows the 0.5 ps pumping pulse, implying that much of the LP relaxation has already taken place  during the pumping stage. Under weak illumination (Fig. \ref{fig:lifetime_all}a), the cavity photon population  can reach 0.01 a.u. ($\approx \hbar\omega_0 = 2327$ cm$^{-1}$), implying that the system receives about one quantum of energy and stays on the singly excited manifold. By contrast, under strong illumination (Fig. \ref{fig:lifetime_all}b), the cavity photon population signals can reach 1.0 a.u. ($\approx 100 \hbar\omega_0$), meaning that the system is very highly excited.

	Because we have assumed no cavity loss and  we find that, outside a cavity, vibrational relaxation of an individual molecule to the ground state takes much longer than several ps, the fast (< 5 ps) polariton relaxation observed in Fig. \ref{fig:lifetime_all} must reflect energy transfer to the vibrational dark modes of the asymmetric \ch{C=O} stretch. Therefore, the fact that the LP relaxation is faster than that of the UP 
	implies either a stronger interaction between the LP and individual molecular motions, or the existence of a decay channel for the LP that is not open for the UP.
	Below we provide evidence in support of the latter scenario. This new decay channel, which we will call polariton-enhanced molecular nonlinear absorption, can exist under both weak and strong illumination of the LP (see Fig. \ref{fig:ratio} for details). Due to this new decay channel, in both Figs. \ref{fig:lifetime_all}a,b, the LP signal exhibits a shortened height and lifetime compared to the UP signal.
	A more detailed study of vibrational polariton relaxation combined with analytical theory and CavMD simulations will be given elsewhere.

		\begin{figure*}
		\centering
		\includegraphics[width=1.0\linewidth]{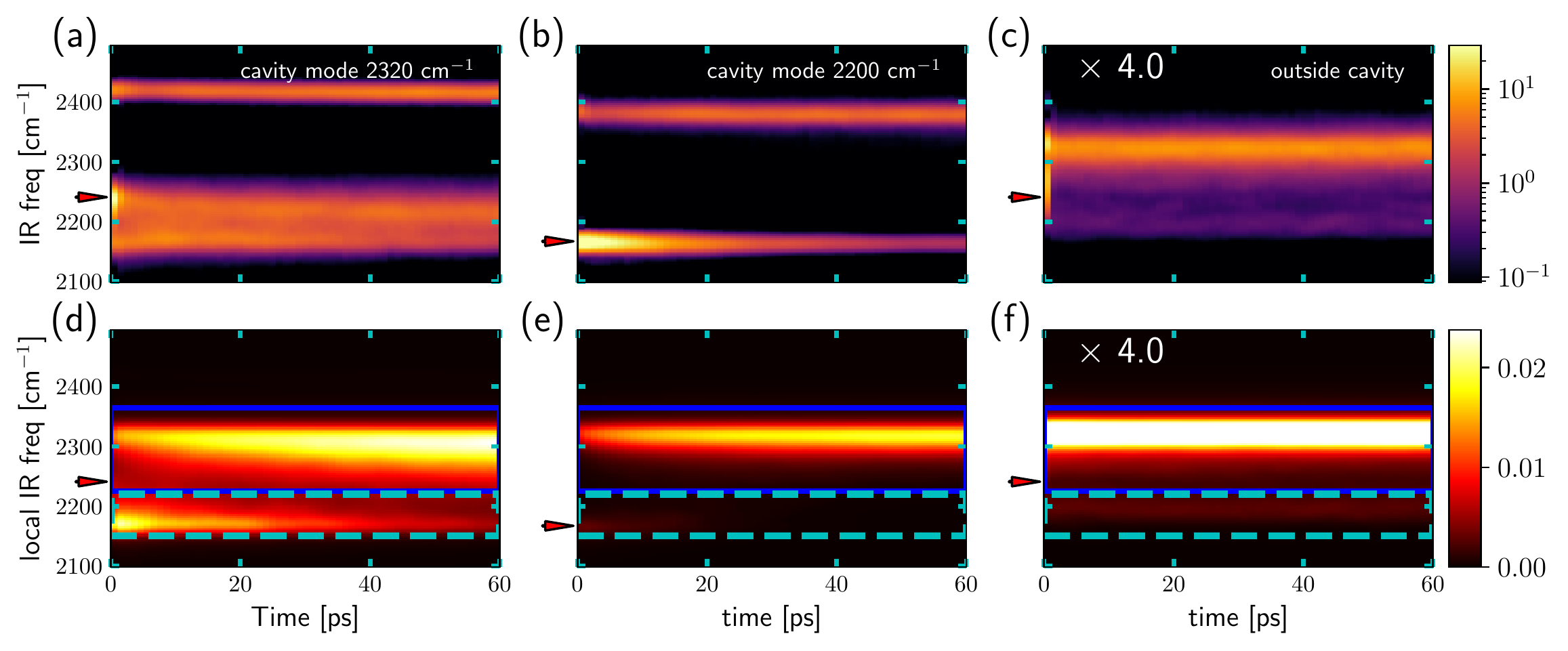}
		\caption{Time-resolved spectra after exciting the LP with a strong pulse ($F = 632$ mJ/cm$^{2}$). 
		Three cases are compared: Figs. a,d: exciting the LP (2241 cm$^{-1}$) inside a  2320 cm$^{-1}$ cavity; Figs. b,e: exciting LP (2167 cm$^{-1}$) inside a 2200 cm$^{-1}$ cavity; Figs. c,f: off-resonant excitation at 2241 cm$^{-1}$ outside the cavity.  The corresponding incident exciting frequency is also labeled as a red arrow in each subplot, and \textit{Figs. c,f have been multiplied by a factor of four for better visualization}. 
		Here, the IR spectra (top panel, evaluated with Eq. \eqref{eq:IR_equation_cavity}) reflects information about the molecular collective bright state, while the local IR spectra	(bottom panel, evaluated with Eq. \eqref{eq:localIR_equation_cavity}) reflects mostly  information about the molecular dark modes.
		At every time snapshot $T_i$, the IR or local IR spectrum is calculated by averaging over the time period $[T_i, T_i + \Delta T]$ with Eq. \eqref{eq:IR_equation_cavity} or \eqref{eq:localIR_equation_cavity}, where $\Delta T = 5$ ps. See Sec. \ref{sec:simu} for other simulation details. 
		To better distinguish between the linear and nonlinear absorption, the region of the linear absorption is labeled within blue horizontal lines (i.e., from 2220 to 2360 cm$^{-1}$) and the region of the nonlinear absorption is labeled within cyan horizontal lines (i.e., from 2150 to 2220 cm$^{-1}$).
		Note that inside the cavity, when exciting the LP, the nonlinear absorption can be greatly enhanced than that outside the cavity; see Figs. \ref{fig:dynamics_CO2_LP}d,f for comparison.
	 	After the pulse, the system temperature is increased from 300 K to 505 K, 366 K, 331 K (from left to right), respectively.
			}
		\label{fig:dynamics_CO2_LP}
	\end{figure*}

	\subsection{Polariton-enhanced nonlinear absorption}\label{sec:result_LP}

	In order to demonstrate that polariton-enhanced molecular nonlinear absorption is the origin of the ultrashort LP lifetime in Fig. \ref{fig:lifetime_all}b, we  next study the molecular response to the pulse excitation as expressed by its transient IR spectrum. Here, the transient IR spectrum  about time $T_i$ is calculated by evaluating Eq. \eqref{eq:IR_equation_cavity} in a time window $[T_i, T_i + \Delta T]$, where $\Delta T = 5$ ps. Note that in this classical calculation the spectrum in Eq. \eqref{eq:IR_equation_cavity} yields information about the molecular frequency distribution, and the corresponding transient spectrum corresponds to this distribution in the excited molecular ensemble.
	
	\subsubsection{Cavity mode at 2320 cm$^{-1}$}
	
	For the same parameters as in Fig. \ref{fig:lifetime_all}b, Fig. \ref{fig:dynamics_CO2_LP}a shows the time-resolved IR spectra after exciting the LP at 2241 cm$^{-1}$ (the red arrow) with a strong pulse ($F=632$ mJ/cm$^{-1}$). Two observations can be made:  (a) There is an ultrafast relaxation of the LP signal which disappears almost immediately after the exciting pulse, and (b)  a distribution of lower frequencies, peaked at $\sim 2170$ cm$^{-1}$, emerges. 
	The existence of such lower frequencies indicates the appearance of molecules with higher vibrational energies (see Appendix \ref{App:freq2vib} where we establish an explicit semiclassical relationship between the vibrational frequency and vibrational quanta for our anharmonic \ch{CO2} force field). In particular, for the anharmonic potential that we use to simulate \ch{CO2} dynamics, the frequency $\sim 2200$ cm$^{-1}$ roughly corresponds to the motion of a classical anharmonic oscillator whose amplitude is determined by having two quanta of vibrational energy.  Quantum mechanically, this frequency corresponds to the energy of the $1\rightarrow 2$ vibrational transition. This suggests that the additional peak $\sim 2170$ cm$^{-1}$ is a signal of nonlinear multiphoton absorption.

	For the same transient state that yields Fig. \ref{fig:dynamics_CO2_LP}a,
	Fig. \ref{fig:dynamics_CO2_LP}d plots the corresponding time-resolved local IR spectra obtained from Eq. \eqref{eq:localIR_equation_cavity} which reflects information about the individual molecular modes (which are mostly composed of vibrational dark modes) rather than a collective bright state. During and after excitation of the LP at 2241 cm$^{-1}$ (red arrow), a large fraction of the energy is transferred to the higher vibrational  excited states of the \ch{C=O} asymmetric stretch, leading to a broad distribution of frequencies and showing a peak at $\sim 2170$ cm$^{-1}$ (see the frequency  region in the cyan rectangle); we also find a peak around 2327 cm$^{-1}$ (see the frequency  region in the blue rectangle) that corresponds to the absorption of individual molecules, where the asymmetric \ch{C=O} stretches are in a thermal distribution. As time increases, the excess energy in  the aforementioned higher excited states gradually relaxes to the latter thermal distribution of vibrational states. This interpretation can be validated by noting that the fitted lifetimes for the energy decay (39 ps) and gain (34 ps) processes are consistent. In other words, due to the polariton-enhanced molecular nonlinear absorption mechanism, the population gain of the \ch{C=O} asymmetric stretches in a  thermal distribution (with predominantly 0 or 1 vibrational quanta) is significantly delayed compared with the ultrafast LP lifetime. 
		
	Interestingly, a recent experiment performed by Xiong \textit{et al} \cite{Xiang2019State} reported 
	that, for \ch{W(CO)6}, after an excitation of the LP, there is a significant delay in the gain of population in the vibrationally first excited state of the dark-state manifold (which corresponds to the  lower vibrational excited states in the blue rectangle of Fig. \ref{fig:dynamics_CO2_LP}d). In Ref. \cite{Xiang2019State}, the authors proposed that the underlying mechanism should be the direct transition from the LP to second or higher vibrational excited states and the subsequent relaxation to the first excited state. In fact, after the submission of this manuscript, via private communication, we have learned \cite{WXiong2020private} of  experimental data which would appear to suggest that, for \ch{W(CO)6}, pumping the LP leads to excitation of a vibrationally second excited state, followed by a delay and then population of a first vibrationally excited state (in agreement with our numerical results above). Note that their results  have a weaker nonlinear signal presumably because the excitation of the LP was not very strong in their work.

	Now, although the experiments in Ref. \cite{Xiang2019State} would appear to validate the  CavMD results in Fig. \ref{fig:dynamics_CO2_LP}d, a keen reader may still be very surprised that a classical simulation can produce a two-peak feature in Fig. \ref{fig:dynamics_CO2_LP}d. 
	Whereas within a quantum model, there are several different vibrational transition energies for an anharmonic oscillator,  it is widely known that a classical response function should predict only a single vibrational peak for a single anharmonic oscillator driven by an external field. 
	
	\begin{figure}
		\centering
		\includegraphics[width=1.0\linewidth]{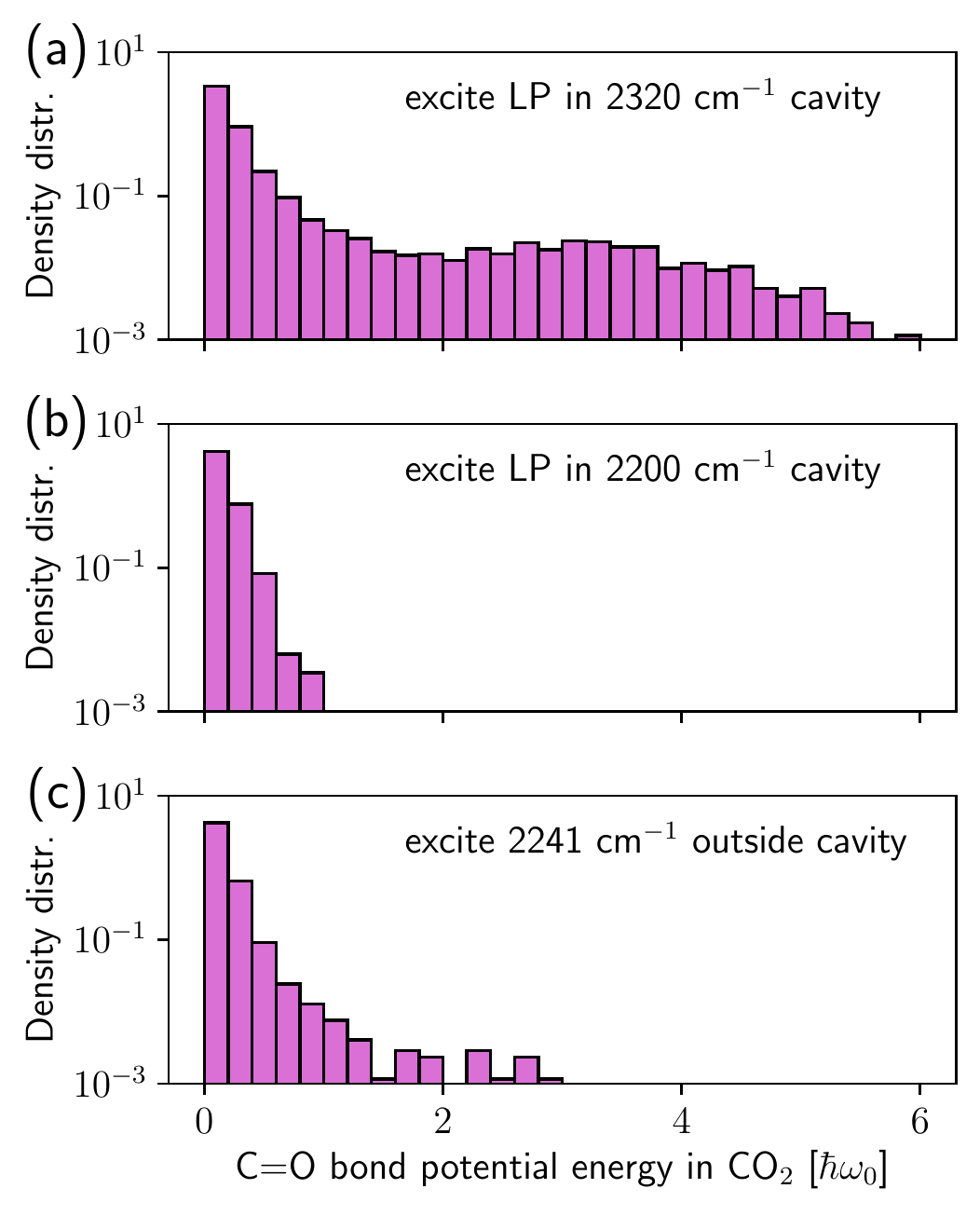}
		\caption{The corresponding density distribution of \ch{C=O} bond potential energy per molecule at time $T_i = 1$ ps for Fig. \ref{fig:dynamics_CO2_LP}. Three cases are compared: exciting the LP inside a (a) 2320 cm$^{-1}$ or (b) 2200 cm$^{-1}$ cavity, and (c) exciting 2241 cm$^{-1}$ (the LP frequency for Fig. a) outside the cavity. All parameters are the same as Fig. \ref{fig:dynamics_CO2_LP}. The \ch{C=O} bond potential energy is calculated according to Eq. \eqref{eq:VCO} and is shown in units of $\hbar\omega_0$ ($\omega_0 = 2327$ cm$^{-1}$). Note that exciting LP in a 2320 cm$^{-1}$ cavity induces a meaningful fraction of vibrationally highly excited molecules, which corresponds to the nonlinear local IR peak $\sim 2170$ cm$^{-1}$ in Fig. \ref{fig:dynamics_CO2_LP}d.
		}
		\label{fig:VCO_energy_dist_LP}
	\end{figure}
	
	In order to understand the origin of the two-peak feature in Fig. \ref{fig:dynamics_CO2_LP}d,  we directly study  the probability density for the \ch{C=O} bond potential energy. Fig. \ref{fig:VCO_energy_dist_LP}a plots this probability distribution (in logarithmic scale) at an early time just after the pulse ($T_i = 1$ ps)  --- under the same conditions as Fig. \ref{fig:dynamics_CO2_LP}d, Fig. \ref{fig:VCO_energy_dist_LP}a clearly shows that the \ch{C=O} vibrational energy probability distribution has not only a strong peak in the low (thermal) energy regime, but also a large  shoulder for states above $2\hbar\omega_0$ ($\omega_0 = 2327$ cm$^{-1}$). As shown in Appendix Fig. \ref{fig:freq2vib}, since a higher vibrational energy corresponds to a lower vibrational frequency within the anharmonic \ch{C=O} force field, the strong peak and the large shoulder in Fig. \ref{fig:VCO_energy_dist_LP}a will lead to two peaks in Fig. \ref{fig:dynamics_CO2_LP}d if we consider an ensemble average of all molecules.
	
	\subsubsection{Excitation outside the cavity}
	The data above suggest that, for some experiments, exciting the system at the LP frequency can facilitate large nonlinear absorption of energy. 
	To confirm that the LP is in fact facilitating such an effect,
	in Figs. \ref{fig:dynamics_CO2_LP}c,f we plot the corresponding IR and local IR spectra when the molecules are  excited outside the cavity with the same strong pulse (centered at 2241 cm$^{-1}$) as in Figs. \ref{fig:dynamics_CO2_LP}a,d. A very weak nonlinear absorption peaked $\sim   2200$ cm$^{-1}$ can be detected; see the dashed cyan region.  As mentioned above, this $\sim   2200$ cm$^{-1}$ weak peak may correspond to a small number of molecules with two or more quanta of vibrational energy. Note that this interpretation agrees with the corresponding \ch{C=O} bond potential energy distribution in Fig. \ref{fig:VCO_energy_dist_LP}c, where we find a very  small fraction of molecules  have \ch{C=O} bond potential energies near $2\hbar\omega_0$, and the large population of highly excited molecules seen in Fig. \ref{fig:dynamics_CO2_LP}d is not reproduced here. 
	Comparing the results inside versus outside the cavity, we conclude that the LP in Figs. \ref{fig:dynamics_CO2_LP}a,d can greatly enhance the multiphoton nonlinear absorption of molecules.
	Note that polariton-enhanced multiphoton absorption under strong illumination has been shown experimentally for various setups, such as organic excitons in a Fabry--P\'erot cavity \cite{Wang2020} and quantum dots near surface plasmons \cite{Fu2012,Rivera2017},  while its   possibility under collective VSC has not been extensively studied \cite{Xiang2019,Xiang2019State, Ribeiro2020} especially under strong illumination. Note also that the reader should not be hesitant in deriving such a conclusion based on the fact that our calculations are entirely classical; classical simulations have long been known to capture molecular
	multiphoton nonlinear absorption outside a cavity\cite{Scandolo1992}.
	
	\subsubsection{Cavity mode at 2200 cm$^{-1}$}

	Finally, let us consider a cavity with a non-resonant frequency mode; such a consideration will lead to a new understanding of the conditions necessary for a polariton to facilitate multiphoton nonlinear absorption. Consider the case when the cavity mode frequency is slightly off resonance with respect to the molecular ground-state frequency. Figs. \ref{fig:dynamics_CO2_LP}b,e show results of similar simulations for a system with cavity mode frequency 2200 cm$^{-1}$.  Now, when the LP (peaked at 2167 cm$^{-1}$) is resonantly excited with a  strong pulse ($F = 632$ mJ/cm$^{-2}$), as shown in Figs. \ref{fig:dynamics_CO2_LP}b,e, the large nonlinear absorption at initial times does not appear; see also Fig. \ref{fig:VCO_energy_dist_LP}b for the corresponding \ch{C=O} bond potential energy distribution.

	Here, readers might be curious why the LP cannot facilitate molecular nonlinear absorption even when the LP frequency 2167 cm$^{-1}$ is very close to the frequency of higher vibrational excited states. We hypothesize that this fact can be best understood within a quantum picture. In a quantum picture, when twice the LP frequency  matches the $0 \rightarrow 2$ vibrational transition,  one can expect an enhancement of molecular nonlinear absorption, so that the LP can function as a "virtual state" for exciting dark modes. By contrast, if the LP frequency matches the $1 \rightarrow 2$ transition (as similar to the present case), the total effect of exciting the LP should be small, since vibrational dark modes obey a thermal distribution and  the $v = 1$ population is small at room temperature. Overall,  as a rule of thumb, according to Fig. \ref{fig:VCO_energy_dist_LP}, it would appear that molecular nonlinear absorption is  facilitated when the LP frequency sits in between 2327 cm$^{-1}$ (the frequency of the \ch{CO2} molecules in a thermal distribution) and 2170 cm$^{-1}$ (the frequency of the highly excited \ch{CO2} molecules).
	
	In the present case, the LP frequency does not match the frequency needed to move the molecule above its first excited state (i.e., the virtual state near 2241 cm$^{-1}$ as mentioned above), so the LP cannot function as a resonant gateway for nonlinear molecular absorption. 
	Consequently, the LP lifetime becomes much longer: An exponential fit of the LP intensity in the time-resolved IR spectra (Fig. \ref{fig:dynamics_CO2_LP}b) gives $\tau_{\text{LP}} = 8.5$ ps, while a consistent result of $\tau_{\text{LP}} = 7.5$ ps can be obtained by an exponential fit of the photonic energy (which is calculated in the same way as in Fig. \ref{fig:lifetime_all}). 
	In this relatively slow relaxation process, energy is directly transferred from the LP to the lower excited states of individual molecules and there is no delay in the population gain of the vibrational singly excited manifold.

	\subsubsection{Pulse intensity dependence of  nonlinear absorption}
	
	\begin{figure}
		\centering
		\includegraphics[width=1.0\linewidth]{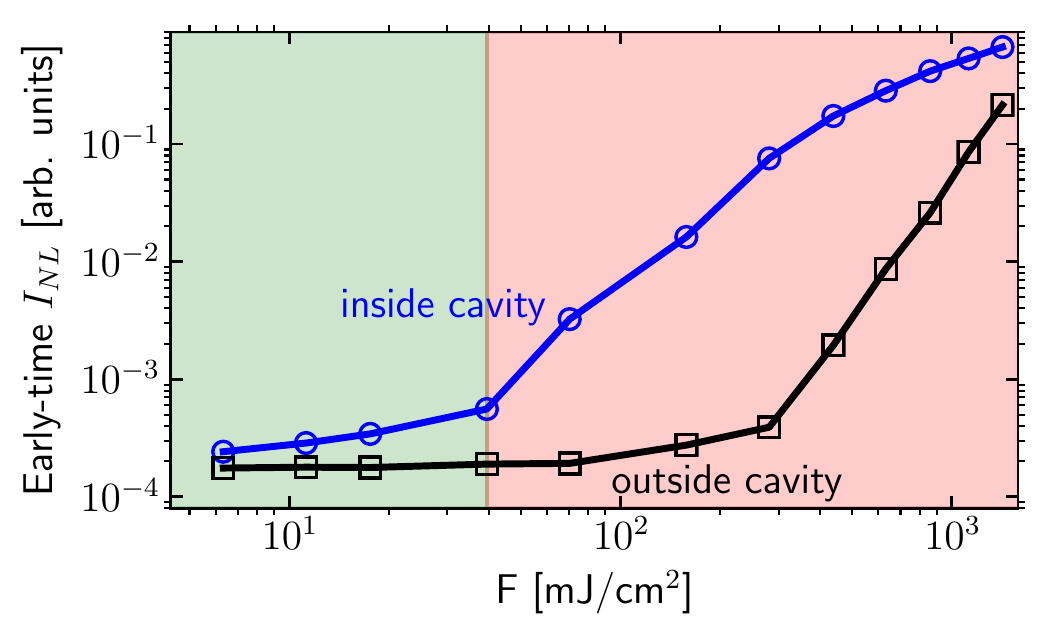}
		\caption{The integrated nonlinear absorption ($I_{\text{NL}}$) at early times plotted as a function of the fluence of the exciting pulse of frequency 2241 cm$^{-1}$ corresponding to the LP frequency in Figs. \ref{fig:dynamics_CO2_LP}a,d. $I_{\text{NL}}$ is calculated by integrating  the local IR spectra over frequency for the nonlinear absorption (i.e., the  cyan region in Fig. \ref{fig:dynamics_CO2_LP}d) at time $T_i =1$  ps. All other parameters are the same as in Figs. \ref{fig:lifetime_all}a,b.  Note that compared with the nonlinear absorption outside the cavity (black squares) under the same pulse, the nonlinear signal inside the cavity (blue circles) can be enhanced by up to two orders of magnitude by polariton-enhanced multiphoton absorption.}
		\label{fig:ratio}
	\end{figure}
	
	The cavity effect in enhancing nonlinear absorption can be further seen by studying the nonlinear dependence on the pulse fluence in and outside the cavity. Fig. \ref{fig:ratio} shows the   local IR spectrum from Fig. \ref{fig:dynamics_CO2_LP}d, integrated over the frequency region that corresponds to the nonlinear absorption inside a 2320 cm$^{-1}$ cavity (the same as Figs. \ref{fig:dynamics_CO2_LP}a,d) and outside a cavity,   as a function of the pulse fluence. 
	In order to make a quantitative analysis, we define the relevant integration region as the area colored cyan in Fig. \ref{fig:dynamics_CO2_LP}d and evaluate this integral at the early time $T_i=1$ ps.
	This early-time nonlinear signal ($I_{\text{NL}}$) provides a direct means to quantify the magnitude of the nonlinearity at the beginning stages of absorption. When the incoming pulse is weak (the green-shadowed region), the nonlinearity is weakly enhanced inside the cavity (blue circles) relative to  the molecular response outside the cavity (black squares). \footnote{Note that this weak enhancement of nonlinearity might be responsible for the relative short LP lifetime as found in Fig. \ref{fig:lifetime_all}a.} By contrast, when the incoming pulse is strong (the red-shadowed  region), $I_{\text{NL}}$ increases at both in and outside the cavity, and under the same pulse fluence, the nonlinearity can be enhanced by up to two orders of magnitude inside the cavity in comparison with the free space case. This provides a direct demonstration of the cavity role in promoting and maintaining multiphoton nonlinear absorption inside an optical cavity.
	
	\subsection{Effect of periodic boundary conditions}\label{sec:result_PBC}
		\begin{figure}
		\centering
		\includegraphics[width=1.0\linewidth]{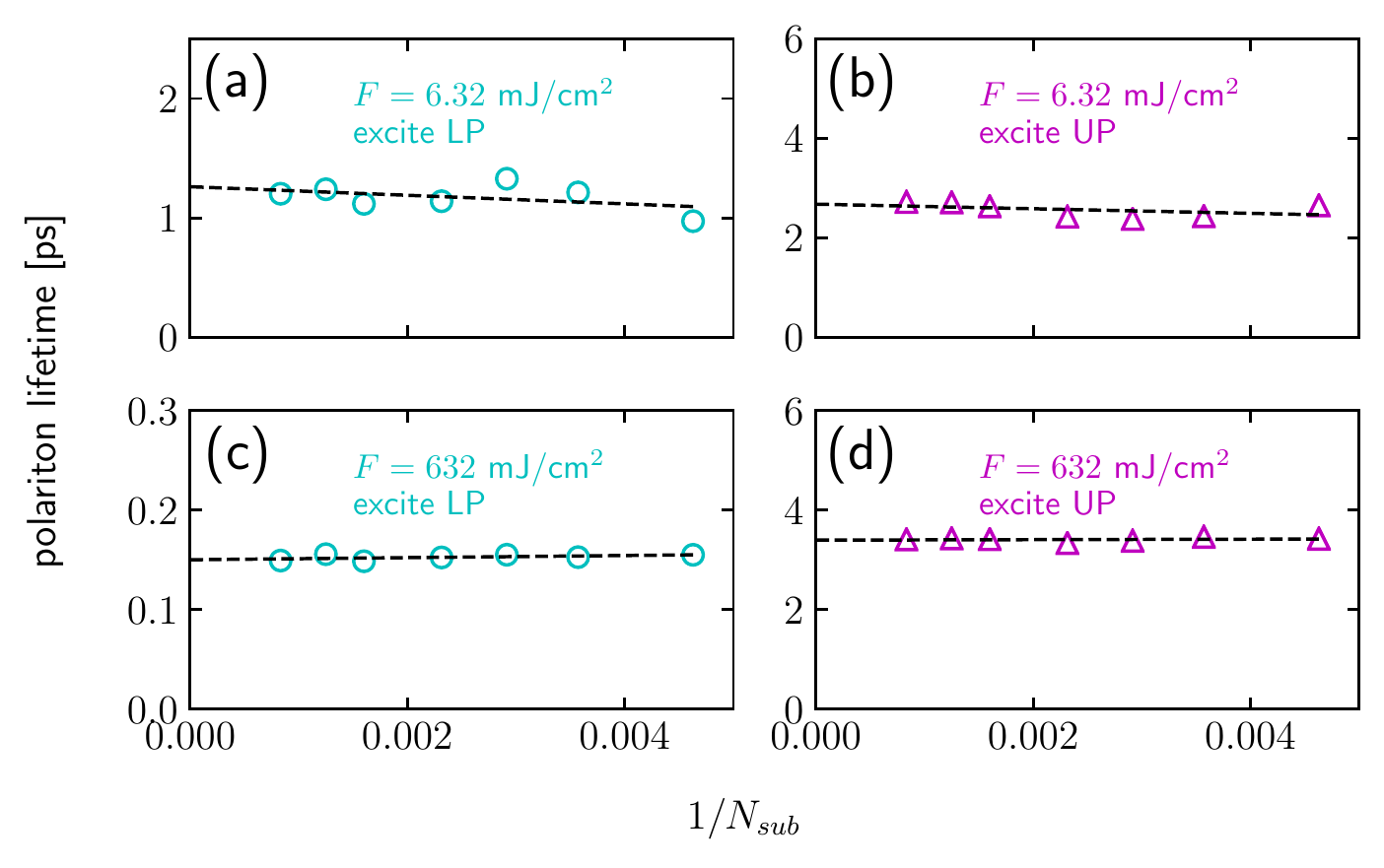}
		\caption{Polariton lifetime against number of molecules in the simulation cell ($1/N_{\text{sub}}$). 
			Figs. a,b: Lifetimes of (a) LP (cyan circles) and (b) UP (magenta triangles) under weak excitation ($F = 6.32$ mJ/cm$^{2}$). Figs. c,d: Lifetimes of (c) LP and (d) UP under strong excitation ($F = 632$ mJ/cm$^{2}$).  Dashed lines denote the linear fitting of each data set. For parameters, when $N_{\text{sub}}$ is increased from 216 to 1200, molecular density and the observed Rabi splitting are fixed the same by increasing the simulation cell size and decreasing the effective coupling strength $\widetilde{\varepsilon}$ accordingly. All other simulation details are the same as Fig. \ref{fig:lifetime_all}.}
		\label{fig:lifetime_number}
	\end{figure}

	Before ending our manuscript, we emphasize that there is a huge numerical gap between realistic VSC experiments and our CavMD simulations. In experiments a macroscopic number of molecules forms VSC in Fabry--P\'erot cavities and  the coupling to a cavity mode for each molecule ($\varepsilon$) is very small. By contrast, in CavMD simulations hundreds of molecules are explicitly simulated and the light-matter interaction per molecule $\widetilde{\varepsilon} = \sqrt{N_{\text{cell}}}\varepsilon$ is artificially amplified proportionally to the number of periodic simulation cells ($N_{\text{cell}}$) due to the invoking of periodic boundary conditions (see Ref. \cite{Li2020Water} for details). In order to validate the intriguing simulation results above, we have performed additional simulations to investigate the effect of  the imposed periodic boundary conditions on the simulation results. When macroscopic observables (such as Rabi splitting and molecular density) are kept the same, we increase the number of molecules in the simulation cell ($N_{\text{sub}}$) and study the polariton lifetime dependence on $N_{\text{sub}}$. Note that fixing Rabi splitting (i.e., fixing $N = N_{\text{sub}} N_{\text{cell}}$) while increasing  $N_{\text{sub}}$ will lead to a decrease of $\widetilde{\varepsilon}$  ($\widetilde{\varepsilon} \propto \sqrt{N_{\text{cell}}} = \sqrt{N/N_{\text{sub}}}$) and therefore the reduction of any artifacts due to the use of periodic boundary conditions. 
	
	Fig. \ref{fig:lifetime_number} shows that the polariton lifetimes are unchanged against $1/N_{\text{sub}}$ under the same conditions as in Fig. \ref{fig:lifetime_all}: the left panel plots the LP lifetime under the  weak (Fig. \ref{fig:lifetime_number}a; $F = 6.32$ mJ/cm$^2$) or strong (Fig. \ref{fig:lifetime_number}c; $F = 632$ mJ/cm$^{2}$) pulse as in Fig. \ref{fig:lifetime_all}; the right panel plots the UP lifetime accordingly. For all situations (including the ultrashort LP lifetime in Fig. \ref{fig:lifetime_number}c), the results show no dependence on $1/N_{\text{sub}}$, suggesting that the reported simulation results are not sensitive to renormalization applied to the light-matter interaction ($\widetilde{\varepsilon}$) and implying that, as long as the observed Rabi splitting is fixed, these predictions may hold in cavities with different volumes, ranging from plasmonic cavities to Fabry--P\'erot cavities, the latter of which are usually used for VSC experiments. 
	
	Of course, because small-volume cavities usually have large cavity losses (while we have ignored cavity losses in our simulations), energy transfer between the polariton and the dark modes may not be as strong for such experiments as in the present results. In Fabry--P\'erot cavities, however, because the cavity loss lifetime ($\sim 5$ ps)  is longer than the polaritonic relaxation lifetimes in Fig. \ref{fig:lifetime_number}, the presented results should not be meaningfully altered even when the cavity loss is considered.

	\section{Conclusion}\label{sec:conclusion}

	To summarize our observations, by studying how molecules respond both individually and collectively after a polariton has been weakly excited, we have found that polariton relaxation to molecular dark modes usually occurs on a timescale of several ps, in agreement with previous experiments \cite{Xiang2018,Xiang2019State}. However, when a strong pulse is applied to the LP in a suitable cavity and the LP energy can support transitions to higher molecular states, the LP lifetime can become ultrashort (0.2 ps) and one can find individual molecules in very highly excited vibrational states.  This so-called multiphoton nonlinear molecular absorption of light can be as large as two orders of magnitude relative to the excitation outside the cavity and arises only in concert with the LP (not the UP) because, for a realistic vibration, the $0 \rightarrow 1$ transition always has a larger frequency than the $1 \rightarrow 2$ transition so that the $0 \rightarrow 2$ transition can approximately match twice the LP frequency.

	Given that this LP relaxation behavior is  robust within our simulation, and especially to the choice against of periodic boundary conditions used in our CavMD simulations, we have every reason to believe that our microscopic simulations presented here will have real macroscopic experimental consequences in cavities with different volumes, and we expect such  intriguing ultrashort LP lifetime and LP-enhanced molecular nonlinear absorption will soon be experimentally verified in usual VSC setups such as Fabry--P\'erot cavities (where a macroscopic number of molecules form VSC).
	As highlighted in Sec. \ref{sec:result_LP}, recent experiments by Xiong \textit{et al} have observed a delayed population gain in the singly excited dark state manifold following an excitation of the LP \cite{Xiang2019State}, which would appear to be a strong endorsement of the current CavMD simulations (while also raising the possibility of even more dramatic findings if the LP is strongly illuminated).

	In the present manuscript, we have investigated VSC using an exclusively classical approach. Although a quantum approach would be ideal for studying light-matter interactions, any brute-force approach is not feasible because VSC phenomena involve a large number of molecules. Moreover, given the large number of vibrationally highly excited states, and the complex interactions between bright and dark modes as caused by short-range intermolecular interactions, no easy simplification of the problem seems feasible. Importantly, the effects discussed in the present manuscript are all classical in nature. Given the excellent agreement between our classical CavMD simulations and VSC relaxation experiments, classical CavMD simulations would appear to be a promising tool to investigate VSC-related phenomena, at least qualitatively. This statement is consistent with previous findings that classical theories can qualitatively capture many intriguing light-matter interaction processes that have classical analogs \cite{Gross1982,Li2018Spontaneous}. Obviously, purely quantum effects such as entanglement will need a quantum treatment.
	Looking forward, the open question remains as to if and how the present CavMD simulations to study the key outstanding and unexplained VSC phenomenon: chemical catalysis.

\begin{appendices}
	
	\setcounter{equation}{0}
	\setcounter{table}{0}
	\renewcommand{\theequation}{S\arabic{equation}}
	\renewcommand{\bibnumfmt}[1]{[S#1]}
	\setcounter{section}{0}
	\renewcommand{\thesection}{S-\Roman{section}}
	
	\section{Potential of \ch{C=O} bond}\label{Appendix:CO}	
	\begin{figure}[h]
		\centering
		\includegraphics[width=1.0\linewidth]{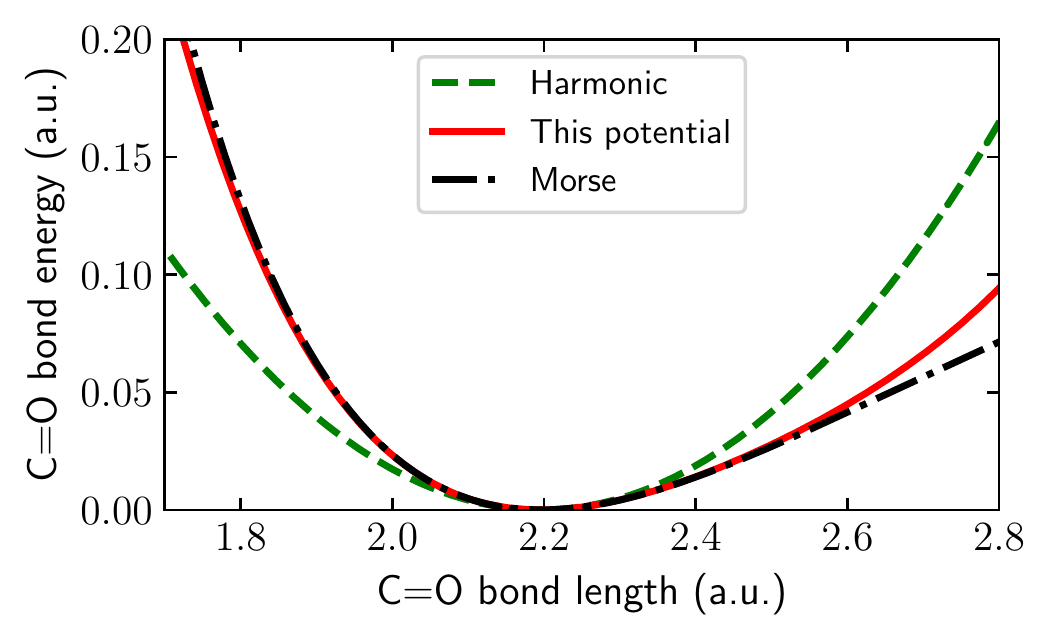}
		\caption{\ch{C=O} bond potential in our force field. The fourth-order anharmonic potential in Eq. \eqref{eq:VCO} (red solid) is compared with the corresponding Morse potential (black dash-dot) and the harmonic approximation (green dashed).}
		\label{fig:anharmonic_potential}
	\end{figure}
	
	The potential of the \ch{C=O} bond in our force field is plotted in Fig. \ref{fig:anharmonic_potential}.

	\section{Correspondence between vibrational frequency and energy}\label{App:freq2vib}

	In simulations we have found the emergence of the low-frequency absorption after exciting the LP. The appearance of low frequencies comes from the use of an anharmonic \ch{C=O} potential as described in Fig. \ref{fig:anharmonic_potential}. Due to the intrinsically anharmonic nature of molecular vibrations, a higher quantum vibrational transition (e.g., $n\rightarrow n+1$ where $n$ is large) has a lower frequency than the fundamental $0\rightarrow 1$ transition. Classically speaking, an anharmonic oscillator also demonstrate a lower, red-shifted frequency when this oscillator contains a higher energy (or known as classical action); see Ref. \cite{Goggin1988} for details. In order to quantify the vibrational state for the corresponding frequencies, we perform a simple simulation for a bare \ch{CO2} in gas phase after an external pulse peaked at 2327 cm$^{-1}$ with duration 0.1 ps; see Eq. \eqref{eq:Ex}. Because this pulse has a shorter duration than that used in the main text (0.5 ps), it contains a wide spectrum band and can excite the molecule to a wide range of vibrational excited states by increasing the pulse fluence. The simulation is performed for 20 ps in a NVE ensemble and the initial configuration of  the molecule is set at the global minimum of potential energy surface with no initial velocity. Under different pulse amplitudes, the \ch{C=O} asymmetric stretch can oscillate with different  vibrational energies.

\begin{figure}
		\centering
		\includegraphics[width=1.0\linewidth]{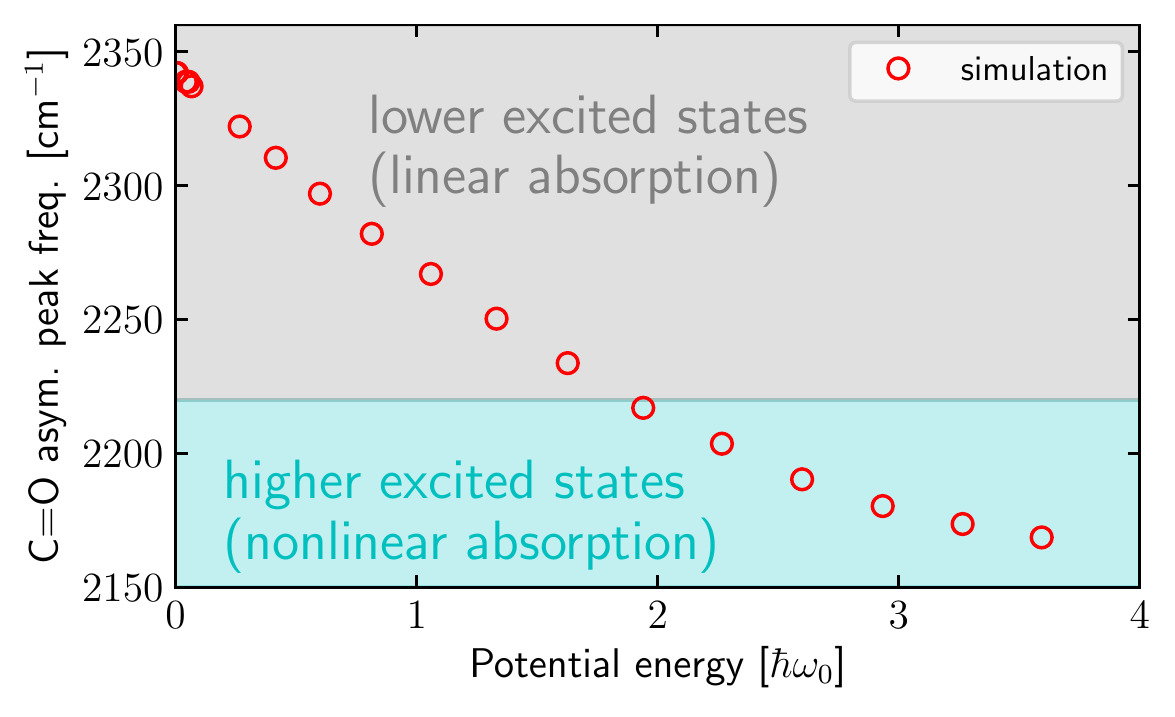}
		\caption{\ch{C=O} asymmetric peak frequency as a function of the potential energy of a single \ch{CO2} molecule in gas phase. 
			The potential energy is plotted in units of $\hbar\omega_0$, where $\omega_0 = 2342$ cm$^{-1}$ denotes the peak frequency in thermal equilibrium at 300 K.	
			See text around for simulation details. }
		\label{fig:freq2vib}
	\end{figure}

	Fig. \ref{fig:freq2vib} plots the corresponding \ch{C=O} asymmetric peak frequency by evaluating the dipole auto-correlation function in Eq. \eqref{eq:IR_equation_cavity} as a function of the vibrational energy of the  molecule, which is predominately contributed by the vibrational energy of the \ch{C=O} asymmetric stretch mode.
	While at thermal equilibrium the \ch{C=O} asymmetric stretch peaks at $\omega_0=2342$ cm$^{-1}$, Fig. \ref{fig:freq2vib} shows that the peak frequency exhibits a negative  relationship with the vibrational potential energy. For example, the frequency near $2200$ cm$^{-1}$ corresponds  to roughly two vibrational quanta. According to the correspondence between frequency  and vibrational energy, we attribute the frequency  range $[2220, 2360)$  cm$^{-1}$ as lower excited states (or linear  absorption; see the blue region)  and   the frequency  range $[2150, 2220)$  cm$^{-1}$ as higher excited states, (or nonlinear  absorption; the cyan region).

	\section{\ch{CO2} force field}\label{App:force_field}
	For the completeness of this manuscript, below we will provide the details of \ch{CO2} force field. This force field largely resembles Ref. \cite{Cygan2012} except for the use of an anharmonic \ch{C=O} bond potential in Eq. \eqref{eq:VCO}.
	
	In this \ch{CO2} force field, the  pairwise intermolecular potential is characterized  by the Lennard--Jones potential   ($V_{\text{LJ}}^{(nl)}$)
	plus the Coulombic interactions between atoms ($V_{\text{Coul}}^{(nl)}$):
	\begin{equation}
		V_{\text{inter}}^{(nl)} = V_{\text{LJ}}^{(nl)} + V_{\text{Coul}}^{(nl)}
	\end{equation}
	where the superscript $n, l$ denote the indices of different molecules. The form of $V_{\text{LJ}}^{(nl)}$ is
	\begin{equation}
		V_{\text{LJ}}^{(nl)} = \sum_{i\in n} \sum_{j \in l} 4\epsilon_{ij} \left[\left(\frac{\sigma_{ij}}{R_{ij}}\right)^{12} - \left(\frac{\sigma_{ij}}{R_{ij}}\right)^{6} \right]
	\end{equation}
	where  $R_{ij}$ ($i \in n$ and $j \in l$) denotes the distance between atoms in molecules $n$ and $l$. For parameters,  $\epsilon_{\ch{CC}} = 0.0559 \text{\ kcal mol}^{-1}$ ($8.9126\times 10^{-5}$ a.u.), $\sigma_{\ch{CC}} = 2.800 \ \si{\angstrom}$ (5.291 a.u.), $\epsilon_{\ch{OO}} = 0.1597 \text{\ kcal mol}^{-1}$ ($2.5454 \times 10^{-4}$ a.u.), $\sigma_{\ch{OO}} = 3.028 \ \si{\angstrom}$ (5.722 a.u.), $\epsilon_{\ch{CO}} = \sqrt{\epsilon_{\ch{CC}}  \epsilon_{\ch{OO}} }$, and $\sigma_{\ch{CO}} = (\sigma_{\ch{CC}} + \sigma_{\ch{OO}})/2$.
 	The form of $V_{\text{Coul}}^{(nl)}$ is
	\begin{equation}
		V_{\text{Coul}}^{(nl)} =  
		\sum_{i\in n} \sum_{j \in l} \frac{Q_i Q_j}{\red{4\pi\epsilon_0}R_{ij}}
	\end{equation}
	 For parameters, $Q_{\ch{C}} = 0.6512\ |e|$, and $Q_{\ch{O}} = -0.3256\ |e|$ (where $e$ denotes the charge of the electron).

	The intramolecular interaction is characterized by
	\begin{equation}
		V_{\text{g}}^{(n)} =  V_{\ch{CO}}(R_{n1}) + V_{\ch{CO}}(R_{n2}) 
		+ \frac{1}{2}k_{\theta} \left(\theta_n - \theta_{\text{eq}}\right)^2
	\end{equation}
	Here, the form of $V_{\ch{CO}}$ has been defined in Eq. \eqref{eq:VCO}. $R_{n1}$ and $R_{n2}$ denote the lengths of two \ch{C=O} bonds, $\theta_n$ 
	and $\theta_\text{eq}$ denote the \ch{O-C-O} angle and the equilibrium angle. For parameters, $k_{\theta} = 108.0 \text{ \ kJ mol}^{-1}\text{rad}^{-2}$ (0.0861 a.u.), and $\theta_{\text{eq}} = 180 \text{ deg}$.
	
	Given the \ch{CO2} force field defined above, one can  easily calculate the cavity-free force $\mathbf{F}_{nj}^{(0)}$ as a function of the nuclear configurations by standard molecular dynamics packages. The molecular dipole moment projected along direction $\vxi_{\lambda}=\ve_x, \ve_y$ ($d_{ng, \lambda}$) is given by
	\begin{equation}
		\begin{aligned}
			d_{ng, \lambda} &=\left[ Q_{\ch{O}}\left(\mathbf{R}_{n\ch{O1}}
			+ \mathbf{R}_{n\ch{O2}}\right) + Q_{\ch{C}}\mathbf{R}_{n\ch{C}}\right] \cdot \vxi_{\lambda}
		\end{aligned}
	\end{equation}
	and the derivative $\partial d_{ng,\lambda}/\partial \mathbf{R}_{nj}$ is straightforward. In calculating the IR spectrum, the total dipole moment $\vmu_S$ projected along direction $\vxi_{\lambda}$ is given by $\vmu_S \cdot \vxi_{\lambda} = \sum_{n=1}^{N_{\text{sub}}} d_{ng,\lambda}$.

	\end{appendices}

\section{Acknowledgements}

	This material is based upon work supported by the U.S. Department of Energy, Office of Science, Office of Basic Energy Sciences under Award Number DE-SC0019397 (J.E.S.);
	and US Department of Energy, Office of Science, Basic Energy Sciences, Chemical Sciences, Geosciences, and Biosciences Division (A.N.).
	This research also used resources of the National Energy Research Scientific Computing Center (NERSC), a U.S. Department of Energy Office of Science User Facility operated under Contract No. DE-AC02-05CH11231.

	\section{Data Availability Statement}
	The data that support the findings  of this study (including source code, input and post-processing scripts, and the corresponding tutorials) are openly available on Github at
	\url{ https://github.com/TaoELi/cavity-md-ipi} \cite{TELi2020Github}.


\begin{thebibliography}{55}%
		\makeatletter
		\providecommand \@ifxundefined [1]{%
			\@ifx{#1\undefined}
		}%
		\providecommand \@ifnum [1]{%
			\ifnum #1\expandafter \@firstoftwo
			\else \expandafter \@secondoftwo
			\fi
		}%
		\providecommand \@ifx [1]{%
			\ifx #1\expandafter \@firstoftwo
			\else \expandafter \@secondoftwo
			\fi
		}%
		\providecommand \natexlab [1]{#1}%
		\providecommand \enquote  [1]{``#1''}%
		\providecommand \bibnamefont  [1]{#1}%
		\providecommand \bibfnamefont [1]{#1}%
		\providecommand \citenamefont [1]{#1}%
		\providecommand \href@noop [0]{\@secondoftwo}%
		\providecommand \href [0]{\begingroup \@sanitize@url \@href}%
		\providecommand \@href[1]{\@@startlink{#1}\@@href}%
		\providecommand \@@href[1]{\endgroup#1\@@endlink}%
		\providecommand \@sanitize@url [0]{\catcode `\\12\catcode `\$12\catcode
			`\&12\catcode `\#12\catcode `\^12\catcode `\_12\catcode `\%12\relax}%
		\providecommand \@@startlink[1]{}%
		\providecommand \@@endlink[0]{}%
		\providecommand \url  [0]{\begingroup\@sanitize@url \@url }%
		\providecommand \@url [1]{\endgroup\@href {#1}{\urlprefix }}%
		\providecommand \urlprefix  [0]{URL }%
		\providecommand \Eprint [0]{\href }%
		\providecommand \doibase [0]{https://doi.org/}%
		\providecommand \selectlanguage [0]{\@gobble}%
		\providecommand \bibinfo  [0]{\@secondoftwo}%
		\providecommand \bibfield  [0]{\@secondoftwo}%
		\providecommand \translation [1]{[#1]}%
		\providecommand \BibitemOpen [0]{}%
		\providecommand \bibitemStop [0]{}%
		\providecommand \bibitemNoStop [0]{.\EOS\space}%
		\providecommand \EOS [0]{\spacefactor3000\relax}%
		\providecommand \BibitemShut  [1]{\csname bibitem#1\endcsname}%
		\let\auto@bib@innerbib\@empty
		\bibitem [{\citenamefont {Herrera}\ and\ \citenamefont
			{Owrutsky}(2020)}]{Herrera2019}%
		\BibitemOpen
		\bibfield  {author} {\bibinfo {author} {\bibfnamefont {F.}~\bibnamefont
				{Herrera}}\ and\ \bibinfo {author} {\bibfnamefont {J.}~\bibnamefont
				{Owrutsky}},\ }\bibfield  {title} {\enquote {\bibinfo {title} {{Molecular
						polaritons for controlling chemistry with quantum optics}},}\ }\href
		{https://doi.org/10.1063/1.5136320} {\bibfield  {journal} {\bibinfo
				{journal} {J. Chem. Phys.}\ }\textbf {\bibinfo {volume} {152}},\ \bibinfo
			{pages} {100902} (\bibinfo {year} {2020})}\BibitemShut {NoStop}%
		\bibitem [{\citenamefont {Shalabney}\ \emph {et~al.}(2015)\citenamefont
			{Shalabney}, \citenamefont {George}, \citenamefont {Hutchison}, \citenamefont
			{Pupillo}, \citenamefont {Genet},\ and\ \citenamefont
			{Ebbesen}}]{Shalabney2015}%
		\BibitemOpen
		\bibfield  {author} {\bibinfo {author} {\bibfnamefont {A.}~\bibnamefont
				{Shalabney}}, \bibinfo {author} {\bibfnamefont {J.}~\bibnamefont {George}},
			\bibinfo {author} {\bibfnamefont {J.}~\bibnamefont {Hutchison}}, \bibinfo
			{author} {\bibfnamefont {G.}~\bibnamefont {Pupillo}}, \bibinfo {author}
			{\bibfnamefont {C.}~\bibnamefont {Genet}},\ and\ \bibinfo {author}
			{\bibfnamefont {T.~W.}\ \bibnamefont {Ebbesen}},\ }\bibfield  {title}
		{\enquote {\bibinfo {title} {{Coherent coupling of molecular resonators with
						a microcavity mode}},}\ }\href {https://doi.org/10.1038/ncomms6981}
		{\bibfield  {journal} {\bibinfo  {journal} {Nat. Commun.}\ }\textbf {\bibinfo
				{volume} {6}},\ \bibinfo {pages} {5981} (\bibinfo {year} {2015})}\BibitemShut
		{NoStop}%
		\bibitem [{\citenamefont {George}\ \emph {et~al.}(2015)\citenamefont {George},
			\citenamefont {Shalabney}, \citenamefont {Hutchison}, \citenamefont {Genet},\
			and\ \citenamefont {Ebbesen}}]{George2015}%
		\BibitemOpen
		\bibfield  {author} {\bibinfo {author} {\bibfnamefont {J.}~\bibnamefont
				{George}}, \bibinfo {author} {\bibfnamefont {A.}~\bibnamefont {Shalabney}},
			\bibinfo {author} {\bibfnamefont {J.~A.}\ \bibnamefont {Hutchison}}, \bibinfo
			{author} {\bibfnamefont {C.}~\bibnamefont {Genet}},\ and\ \bibinfo {author}
			{\bibfnamefont {T.~W.}\ \bibnamefont {Ebbesen}},\ }\bibfield  {title}
		{\enquote {\bibinfo {title} {{Liquid-Phase Vibrational Strong Coupling}},}\
		}\href {https://doi.org/10.1021/acs.jpclett.5b00204} {\bibfield  {journal}
			{\bibinfo  {journal} {J. Phys. Chem. Lett.}\ }\textbf {\bibinfo {volume}
				{6}},\ \bibinfo {pages} {1027--1031} (\bibinfo {year} {2015})}\BibitemShut
		{NoStop}%
		\bibitem [{\citenamefont {George}\ \emph {et~al.}(2016)\citenamefont {George},
			\citenamefont {Chervy}, \citenamefont {Shalabney}, \citenamefont {Devaux},
			\citenamefont {Hiura}, \citenamefont {Genet},\ and\ \citenamefont
			{Ebbesen}}]{George2016}%
		\BibitemOpen
		\bibfield  {author} {\bibinfo {author} {\bibfnamefont {J.}~\bibnamefont
				{George}}, \bibinfo {author} {\bibfnamefont {T.}~\bibnamefont {Chervy}},
			\bibinfo {author} {\bibfnamefont {A.}~\bibnamefont {Shalabney}}, \bibinfo
			{author} {\bibfnamefont {E.}~\bibnamefont {Devaux}}, \bibinfo {author}
			{\bibfnamefont {H.}~\bibnamefont {Hiura}}, \bibinfo {author} {\bibfnamefont
				{C.}~\bibnamefont {Genet}},\ and\ \bibinfo {author} {\bibfnamefont {T.~W.}\
				\bibnamefont {Ebbesen}},\ }\bibfield  {title} {\enquote {\bibinfo {title}
				{{Multiple Rabi Splittings under Ultrastrong Vibrational Coupling}},}\ }\href
		{https://doi.org/10.1103/PhysRevLett.117.153601} {\bibfield  {journal}
			{\bibinfo  {journal} {Phys. Rev. Lett.}\ }\textbf {\bibinfo {volume} {117}},\
			\bibinfo {pages} {153601} (\bibinfo {year} {2016})}\BibitemShut {NoStop}%
		\bibitem [{\citenamefont {Thomas}\ \emph {et~al.}(2016)\citenamefont {Thomas},
			\citenamefont {George}, \citenamefont {Shalabney}, \citenamefont {Dryzhakov},
			\citenamefont {Varma}, \citenamefont {Moran}, \citenamefont {Chervy},
			\citenamefont {Zhong}, \citenamefont {Devaux}, \citenamefont {Genet},
			\citenamefont {Hutchison},\ and\ \citenamefont {Ebbesen}}]{Thomas2016}%
		\BibitemOpen
		\bibfield  {author} {\bibinfo {author} {\bibfnamefont {A.}~\bibnamefont
				{Thomas}}, \bibinfo {author} {\bibfnamefont {J.}~\bibnamefont {George}},
			\bibinfo {author} {\bibfnamefont {A.}~\bibnamefont {Shalabney}}, \bibinfo
			{author} {\bibfnamefont {M.}~\bibnamefont {Dryzhakov}}, \bibinfo {author}
			{\bibfnamefont {S.~J.}\ \bibnamefont {Varma}}, \bibinfo {author}
			{\bibfnamefont {J.}~\bibnamefont {Moran}}, \bibinfo {author} {\bibfnamefont
				{T.}~\bibnamefont {Chervy}}, \bibinfo {author} {\bibfnamefont
				{X.}~\bibnamefont {Zhong}}, \bibinfo {author} {\bibfnamefont
				{E.}~\bibnamefont {Devaux}}, \bibinfo {author} {\bibfnamefont
				{C.}~\bibnamefont {Genet}}, \bibinfo {author} {\bibfnamefont {J.~A.}\
				\bibnamefont {Hutchison}},\ and\ \bibinfo {author} {\bibfnamefont {T.~W.}\
				\bibnamefont {Ebbesen}},\ }\bibfield  {title} {\enquote {\bibinfo {title}
				{{Ground-State Chemical Reactivity under Vibrational Coupling to the Vacuum
						Electromagnetic Field}},}\ }\href {https://doi.org/10.1002/anie.201605504}
		{\bibfield  {journal} {\bibinfo  {journal} {Angew. Chemie Int. Ed.}\ }\textbf
			{\bibinfo {volume} {55}},\ \bibinfo {pages} {11462--11466} (\bibinfo {year}
			{2016})}\BibitemShut {NoStop}%
		\bibitem [{\citenamefont {Vergauwe}\ \emph {et~al.}(2019)\citenamefont
			{Vergauwe}, \citenamefont {Thomas}, \citenamefont {Nagarajan}, \citenamefont
			{Shalabney}, \citenamefont {George}, \citenamefont {Chervy}, \citenamefont
			{Seidel}, \citenamefont {Devaux}, \citenamefont {Torbeev},\ and\
			\citenamefont {Ebbesen}}]{Vergauwe2019}%
		\BibitemOpen
		\bibfield  {author} {\bibinfo {author} {\bibfnamefont {R.~M.~A.}\
				\bibnamefont {Vergauwe}}, \bibinfo {author} {\bibfnamefont {A.}~\bibnamefont
				{Thomas}}, \bibinfo {author} {\bibfnamefont {K.}~\bibnamefont {Nagarajan}},
			\bibinfo {author} {\bibfnamefont {A.}~\bibnamefont {Shalabney}}, \bibinfo
			{author} {\bibfnamefont {J.}~\bibnamefont {George}}, \bibinfo {author}
			{\bibfnamefont {T.}~\bibnamefont {Chervy}}, \bibinfo {author} {\bibfnamefont
				{M.}~\bibnamefont {Seidel}}, \bibinfo {author} {\bibfnamefont
				{E.}~\bibnamefont {Devaux}}, \bibinfo {author} {\bibfnamefont
				{V.}~\bibnamefont {Torbeev}},\ and\ \bibinfo {author} {\bibfnamefont {T.~W.}\
				\bibnamefont {Ebbesen}},\ }\bibfield  {title} {\enquote {\bibinfo {title}
				{{Modification of Enzyme Activity by Vibrational Strong Coupling of
						Water}},}\ }\href {https://doi.org/10.1002/anie.201908876} {\bibfield
			{journal} {\bibinfo  {journal} {Angew. Chemie Int. Ed.}\ }\textbf {\bibinfo
				{volume} {58}},\ \bibinfo {pages} {15324--15328} (\bibinfo {year}
			{2019})}\BibitemShut {NoStop}%
		\bibitem [{\citenamefont {Thomas}\ \emph {et~al.}(2019)\citenamefont {Thomas},
			\citenamefont {Lethuillier-Karl}, \citenamefont {Nagarajan}, \citenamefont
			{Vergauwe}, \citenamefont {George}, \citenamefont {Chervy}, \citenamefont
			{Shalabney}, \citenamefont {Devaux}, \citenamefont {Genet}, \citenamefont
			{Moran},\ and\ \citenamefont {Ebbesen}}]{Thomas2019_science}%
		\BibitemOpen
		\bibfield  {author} {\bibinfo {author} {\bibfnamefont {A.}~\bibnamefont
				{Thomas}}, \bibinfo {author} {\bibfnamefont {L.}~\bibnamefont
				{Lethuillier-Karl}}, \bibinfo {author} {\bibfnamefont {K.}~\bibnamefont
				{Nagarajan}}, \bibinfo {author} {\bibfnamefont {R.~M.~A.}\ \bibnamefont
				{Vergauwe}}, \bibinfo {author} {\bibfnamefont {J.}~\bibnamefont {George}},
			\bibinfo {author} {\bibfnamefont {T.}~\bibnamefont {Chervy}}, \bibinfo
			{author} {\bibfnamefont {A.}~\bibnamefont {Shalabney}}, \bibinfo {author}
			{\bibfnamefont {E.}~\bibnamefont {Devaux}}, \bibinfo {author} {\bibfnamefont
				{C.}~\bibnamefont {Genet}}, \bibinfo {author} {\bibfnamefont
				{J.}~\bibnamefont {Moran}},\ and\ \bibinfo {author} {\bibfnamefont {T.~W.}\
				\bibnamefont {Ebbesen}},\ }\bibfield  {title} {\enquote {\bibinfo {title}
				{{Tilting a ground-state reactivity landscape by vibrational strong
						coupling}},}\ }\href {https://doi.org/10.1126/science.aau7742} {\bibfield
			{journal} {\bibinfo  {journal} {Science}\ }\textbf {\bibinfo {volume}
				{363}},\ \bibinfo {pages} {615--619} (\bibinfo {year} {2019})}\BibitemShut
		{NoStop}%
		\bibitem [{\citenamefont {Pang}\ \emph {et~al.}(2020)\citenamefont {Pang},
			\citenamefont {Thomas}, \citenamefont {Nagarajan}, \citenamefont {Vergauwe},
			\citenamefont {Joseph}, \citenamefont {Patrahau}, \citenamefont {Wang},
			\citenamefont {Genet},\ and\ \citenamefont {Ebbesen}}]{Pang2020}%
		\BibitemOpen
		\bibfield  {author} {\bibinfo {author} {\bibfnamefont {Y.}~\bibnamefont
				{Pang}}, \bibinfo {author} {\bibfnamefont {A.}~\bibnamefont {Thomas}},
			\bibinfo {author} {\bibfnamefont {K.}~\bibnamefont {Nagarajan}}, \bibinfo
			{author} {\bibfnamefont {R.~M.~A.}\ \bibnamefont {Vergauwe}}, \bibinfo
			{author} {\bibfnamefont {K.}~\bibnamefont {Joseph}}, \bibinfo {author}
			{\bibfnamefont {B.}~\bibnamefont {Patrahau}}, \bibinfo {author}
			{\bibfnamefont {K.}~\bibnamefont {Wang}}, \bibinfo {author} {\bibfnamefont
				{C.}~\bibnamefont {Genet}},\ and\ \bibinfo {author} {\bibfnamefont {T.~W.}\
				\bibnamefont {Ebbesen}},\ }\bibfield  {title} {\enquote {\bibinfo {title}
				{{On the Role of Symmetry in Vibrational Strong Coupling: The Case of
						Charge‐Transfer Complexation}},}\ }\href
		{https://doi.org/10.1002/anie.202002527} {\bibfield  {journal} {\bibinfo
				{journal} {Angew. Chemie Int. Ed.}\ }\textbf {\bibinfo {volume} {59}},\
			\bibinfo {pages} {10436--10440} (\bibinfo {year} {2020})}\BibitemShut
		{NoStop}%
		\bibitem [{\citenamefont {Xiang}\ \emph
			{et~al.}(2019{\natexlab{a}})\citenamefont {Xiang}, \citenamefont {Ribeiro},
			\citenamefont {Li}, \citenamefont {Dunkelberger}, \citenamefont {Simpkins},
			\citenamefont {Yuen-Zhou},\ and\ \citenamefont {Xiong}}]{Xiang2019Nonlinear}%
		\BibitemOpen
		\bibfield  {author} {\bibinfo {author} {\bibfnamefont {B.}~\bibnamefont
				{Xiang}}, \bibinfo {author} {\bibfnamefont {R.~F.}\ \bibnamefont {Ribeiro}},
			\bibinfo {author} {\bibfnamefont {Y.}~\bibnamefont {Li}}, \bibinfo {author}
			{\bibfnamefont {A.~D.}\ \bibnamefont {Dunkelberger}}, \bibinfo {author}
			{\bibfnamefont {B.~B.}\ \bibnamefont {Simpkins}}, \bibinfo {author}
			{\bibfnamefont {J.}~\bibnamefont {Yuen-Zhou}},\ and\ \bibinfo {author}
			{\bibfnamefont {W.}~\bibnamefont {Xiong}},\ }\bibfield  {title} {\enquote
			{\bibinfo {title} {{Manipulating optical nonlinearities of molecular
						polaritons by delocalization}},}\ }\href
		{https://doi.org/10.1126/sciadv.aax5196} {\bibfield  {journal} {\bibinfo
				{journal} {Sci. Adv.}\ }\textbf {\bibinfo {volume} {5}},\ \bibinfo {pages}
			{eaax5196} (\bibinfo {year} {2019}{\natexlab{a}})},\ \Eprint
		{https://arxiv.org/abs/1901.05526} {arXiv:1901.05526} \BibitemShut {NoStop}%
		\bibitem [{\citenamefont {{F. Ribeiro}}\ \emph {et~al.}(2018)\citenamefont {{F.
					Ribeiro}}, \citenamefont {Dunkelberger}, \citenamefont {Xiang}, \citenamefont
			{Xiong}, \citenamefont {Simpkins}, \citenamefont {Owrutsky},\ and\
			\citenamefont {Yuen-Zhou}}]{F.Ribeiro2018}%
		\BibitemOpen
		\bibfield  {author} {\bibinfo {author} {\bibfnamefont {R.}~\bibnamefont {{F.
						Ribeiro}}}, \bibinfo {author} {\bibfnamefont {A.~D.}\ \bibnamefont
				{Dunkelberger}}, \bibinfo {author} {\bibfnamefont {B.}~\bibnamefont {Xiang}},
			\bibinfo {author} {\bibfnamefont {W.}~\bibnamefont {Xiong}}, \bibinfo
			{author} {\bibfnamefont {B.~S.}\ \bibnamefont {Simpkins}}, \bibinfo {author}
			{\bibfnamefont {J.~C.}\ \bibnamefont {Owrutsky}},\ and\ \bibinfo {author}
			{\bibfnamefont {J.}~\bibnamefont {Yuen-Zhou}},\ }\bibfield  {title} {\enquote
			{\bibinfo {title} {{Theory for Nonlinear Spectroscopy of Vibrational
						Polaritons}},}\ }\href {https://doi.org/10.1021/acs.jpclett.8b01176}
		{\bibfield  {journal} {\bibinfo  {journal} {J. Phys. Chem. Lett.}\ }\textbf
			{\bibinfo {volume} {9}},\ \bibinfo {pages} {3766--3771} (\bibinfo {year}
			{2018})}\BibitemShut {NoStop}%
		\bibitem [{\citenamefont {Xiang}\ \emph {et~al.}(2020)\citenamefont {Xiang},
			\citenamefont {Ribeiro}, \citenamefont {Du}, \citenamefont {Chen},
			\citenamefont {Yang}, \citenamefont {Wang}, \citenamefont {Yuen-Zhou},\ and\
			\citenamefont {Xiong}}]{Xiang2020Science}%
		\BibitemOpen
		\bibfield  {author} {\bibinfo {author} {\bibfnamefont {B.}~\bibnamefont
				{Xiang}}, \bibinfo {author} {\bibfnamefont {R.~F.}\ \bibnamefont {Ribeiro}},
			\bibinfo {author} {\bibfnamefont {M.}~\bibnamefont {Du}}, \bibinfo {author}
			{\bibfnamefont {L.}~\bibnamefont {Chen}}, \bibinfo {author} {\bibfnamefont
				{Z.}~\bibnamefont {Yang}}, \bibinfo {author} {\bibfnamefont {J.}~\bibnamefont
				{Wang}}, \bibinfo {author} {\bibfnamefont {J.}~\bibnamefont {Yuen-Zhou}},\
			and\ \bibinfo {author} {\bibfnamefont {W.}~\bibnamefont {Xiong}},\ }\bibfield
		{title} {\enquote {\bibinfo {title} {{Intermolecular vibrational energy
						transfer enabled by microcavity strong light–matter coupling}},}\ }\href
		{https://doi.org/10.1126/science.aba3544} {\bibfield  {journal} {\bibinfo
				{journal} {Science (80-. ).}\ }\textbf {\bibinfo {volume} {368}},\ \bibinfo
			{pages} {665--667} (\bibinfo {year} {2020})}\BibitemShut {NoStop}%
		\bibitem [{\citenamefont {Rudin}\ and\ \citenamefont
			{Reinecke}(1999)}]{Rudin1999}%
		\BibitemOpen
		\bibfield  {author} {\bibinfo {author} {\bibfnamefont {S.}~\bibnamefont
				{Rudin}}\ and\ \bibinfo {author} {\bibfnamefont {T.~L.}\ \bibnamefont
				{Reinecke}},\ }\bibfield  {title} {\enquote {\bibinfo {title} {{Oscillator
						model for vacuum Rabi splitting in microcavities}},}\ }\href
		{https://doi.org/10.1103/PhysRevB.59.10227} {\bibfield  {journal} {\bibinfo
				{journal} {Phys. Rev. B}\ }\textbf {\bibinfo {volume} {59}},\ \bibinfo
			{pages} {10227--10233} (\bibinfo {year} {1999})}\BibitemShut {NoStop}%
		\bibitem [{\citenamefont {Galego}\ \emph {et~al.}(2019)\citenamefont {Galego},
			\citenamefont {Climent}, \citenamefont {Garcia-Vidal},\ and\ \citenamefont
			{Feist}}]{Galego2019}%
		\BibitemOpen
		\bibfield  {author} {\bibinfo {author} {\bibfnamefont {J.}~\bibnamefont
				{Galego}}, \bibinfo {author} {\bibfnamefont {C.}~\bibnamefont {Climent}},
			\bibinfo {author} {\bibfnamefont {F.~J.}\ \bibnamefont {Garcia-Vidal}},\ and\
			\bibinfo {author} {\bibfnamefont {J.}~\bibnamefont {Feist}},\ }\bibfield
		{title} {\enquote {\bibinfo {title} {{Cavity Casimir-Polder Forces and Their
						Effects in Ground-State Chemical Reactivity}},}\ }\href
		{https://doi.org/10.1103/PhysRevX.9.021057} {\bibfield  {journal} {\bibinfo
				{journal} {Phys. Rev. X}\ }\textbf {\bibinfo {volume} {9}},\ \bibinfo {pages}
			{021057} (\bibinfo {year} {2019})}\BibitemShut {NoStop}%
		\bibitem [{\citenamefont {Li}, \citenamefont {Nitzan},\ and\ \citenamefont
			{Subotnik}(2020)}]{Li2020Origin}%
		\BibitemOpen
		\bibfield  {author} {\bibinfo {author} {\bibfnamefont {T.~E.}\ \bibnamefont
				{Li}}, \bibinfo {author} {\bibfnamefont {A.}~\bibnamefont {Nitzan}},\ and\
			\bibinfo {author} {\bibfnamefont {J.~E.}\ \bibnamefont {Subotnik}},\
		}\bibfield  {title} {\enquote {\bibinfo {title} {{On the origin of
						ground-state vacuum-field catalysis: Equilibrium consideration}},}\ }\href
		{https://doi.org/10.1063/5.0006472} {\bibfield  {journal} {\bibinfo
				{journal} {J. Chem. Phys.}\ }\textbf {\bibinfo {volume} {152}},\ \bibinfo
			{pages} {234107} (\bibinfo {year} {2020})},\ \Eprint
		{https://arxiv.org/abs/2002.09977} {arXiv:2002.09977} \BibitemShut {NoStop}%
		\bibitem [{\citenamefont {Campos-Gonzalez-Angulo}\ and\ \citenamefont
			{Yuen-Zhou}(2020)}]{Campos-Gonzalez-Angulo2020}%
		\BibitemOpen
		\bibfield  {author} {\bibinfo {author} {\bibfnamefont {J.~A.}\ \bibnamefont
				{Campos-Gonzalez-Angulo}}\ and\ \bibinfo {author} {\bibfnamefont
				{J.}~\bibnamefont {Yuen-Zhou}},\ }\bibfield  {title} {\enquote {\bibinfo
				{title} {{Polaritonic normal modes in transition state theory}},}\ }\href
		{https://doi.org/10.1063/5.0007547} {\bibfield  {journal} {\bibinfo
				{journal} {J. Chem. Phys.}\ }\textbf {\bibinfo {volume} {152}},\ \bibinfo
			{pages} {161101} (\bibinfo {year} {2020})}\BibitemShut {NoStop}%
		\bibitem [{\citenamefont {Zhdanov}(2020)}]{Zhdanov2020}%
		\BibitemOpen
		\bibfield  {author} {\bibinfo {author} {\bibfnamefont {V.~P.}\ \bibnamefont
				{Zhdanov}},\ }\bibfield  {title} {\enquote {\bibinfo {title} {{Vacuum field
						in a cavity, light-mediated vibrational coupling, and chemical
						reactivity}},}\ }\href {https://doi.org/10.1016/j.chemphys.2020.110767}
		{\bibfield  {journal} {\bibinfo  {journal} {Chem. Phys.}\ }\textbf {\bibinfo
				{volume} {535}},\ \bibinfo {pages} {110767} (\bibinfo {year}
			{2020})}\BibitemShut {NoStop}%
		\bibitem [{\citenamefont {Scholes}, \citenamefont {DelPo},\ and\ \citenamefont
			{Kudisch}(2020)}]{Scholes2020}%
		\BibitemOpen
		\bibfield  {author} {\bibinfo {author} {\bibfnamefont {G.~D.}\ \bibnamefont
				{Scholes}}, \bibinfo {author} {\bibfnamefont {C.~A.}\ \bibnamefont {DelPo}},\
			and\ \bibinfo {author} {\bibfnamefont {B.}~\bibnamefont {Kudisch}},\
		}\bibfield  {title} {\enquote {\bibinfo {title} {{Entropy Reorders Polariton
						States}},}\ }\href {https://doi.org/10.1021/acs.jpclett.0c02000} {\bibfield
			{journal} {\bibinfo  {journal} {J. Phys. Chem. Lett.}\ }\textbf {\bibinfo
				{volume} {11}},\ \bibinfo {pages} {6389--6395} (\bibinfo {year}
			{2020})}\BibitemShut {NoStop}%
		\bibitem [{\citenamefont {Xiang}\ \emph {et~al.}(2018)\citenamefont {Xiang},
			\citenamefont {Ribeiro}, \citenamefont {Dunkelberger}, \citenamefont {Wang},
			\citenamefont {Li}, \citenamefont {Simpkins}, \citenamefont {Owrutsky},
			\citenamefont {Yuen-Zhou},\ and\ \citenamefont {Xiong}}]{Xiang2018}%
		\BibitemOpen
		\bibfield  {author} {\bibinfo {author} {\bibfnamefont {B.}~\bibnamefont
				{Xiang}}, \bibinfo {author} {\bibfnamefont {R.~F.}\ \bibnamefont {Ribeiro}},
			\bibinfo {author} {\bibfnamefont {A.~D.}\ \bibnamefont {Dunkelberger}},
			\bibinfo {author} {\bibfnamefont {J.}~\bibnamefont {Wang}}, \bibinfo {author}
			{\bibfnamefont {Y.}~\bibnamefont {Li}}, \bibinfo {author} {\bibfnamefont
				{B.~S.}\ \bibnamefont {Simpkins}}, \bibinfo {author} {\bibfnamefont {J.~C.}\
				\bibnamefont {Owrutsky}}, \bibinfo {author} {\bibfnamefont {J.}~\bibnamefont
				{Yuen-Zhou}},\ and\ \bibinfo {author} {\bibfnamefont {W.}~\bibnamefont
				{Xiong}},\ }\bibfield  {title} {\enquote {\bibinfo {title} {{Two-dimensional
						infrared spectroscopy of vibrational polaritons}},}\ }\href
		{https://doi.org/10.1073/pnas.1722063115} {\bibfield  {journal} {\bibinfo
				{journal} {Proc. Natl. Acad. Sci.}\ }\textbf {\bibinfo {volume} {115}},\
			\bibinfo {pages} {4845--4850} (\bibinfo {year} {2018})}\BibitemShut {NoStop}%
		\bibitem [{\citenamefont {Xiang}\ \emph
			{et~al.}(2019{\natexlab{b}})\citenamefont {Xiang}, \citenamefont {Ribeiro},
			\citenamefont {Chen}, \citenamefont {Wang}, \citenamefont {Du}, \citenamefont
			{Yuen-Zhou},\ and\ \citenamefont {Xiong}}]{Xiang2019State}%
		\BibitemOpen
		\bibfield  {author} {\bibinfo {author} {\bibfnamefont {B.}~\bibnamefont
				{Xiang}}, \bibinfo {author} {\bibfnamefont {R.~F.}\ \bibnamefont {Ribeiro}},
			\bibinfo {author} {\bibfnamefont {L.}~\bibnamefont {Chen}}, \bibinfo {author}
			{\bibfnamefont {J.}~\bibnamefont {Wang}}, \bibinfo {author} {\bibfnamefont
				{M.}~\bibnamefont {Du}}, \bibinfo {author} {\bibfnamefont {J.}~\bibnamefont
				{Yuen-Zhou}},\ and\ \bibinfo {author} {\bibfnamefont {W.}~\bibnamefont
				{Xiong}},\ }\bibfield  {title} {\enquote {\bibinfo {title} {{State-Selective
						Polariton to Dark State Relaxation Dynamics}},}\ }\href
		{https://doi.org/10.1021/acs.jpca.9b04601} {\bibfield  {journal} {\bibinfo
				{journal} {J. Phys. Chem. A}\ }\textbf {\bibinfo {volume} {123}},\ \bibinfo
			{pages} {5918--5927} (\bibinfo {year} {2019}{\natexlab{b}})}\BibitemShut
		{NoStop}%
		\bibitem [{\citenamefont {Grafton}\ \emph {et~al.}(2020)\citenamefont
			{Grafton}, \citenamefont {Dunkelberger}, \citenamefont {Simpkins},
			\citenamefont {Triana}, \citenamefont {Hernandez}, \citenamefont {Herrera},\
			and\ \citenamefont {Owrutsky}}]{Grafton2020}%
		\BibitemOpen
		\bibfield  {author} {\bibinfo {author} {\bibfnamefont {A.~B.}\ \bibnamefont
				{Grafton}}, \bibinfo {author} {\bibfnamefont {A.~D.}\ \bibnamefont
				{Dunkelberger}}, \bibinfo {author} {\bibfnamefont {B.~S.}\ \bibnamefont
				{Simpkins}}, \bibinfo {author} {\bibfnamefont {J.~F.}\ \bibnamefont
				{Triana}}, \bibinfo {author} {\bibfnamefont {F.~J.}\ \bibnamefont
				{Hernandez}}, \bibinfo {author} {\bibfnamefont {F.}~\bibnamefont {Herrera}},\
			and\ \bibinfo {author} {\bibfnamefont {J.}~\bibnamefont {Owrutsky}},\
		}\bibfield  {title} {\enquote {\bibinfo {title} {{Excited-State
						Vibration-Polariton Transitions and Dynamics in Nitroprusside}},}\ }\href
		{https://doi.org/10.26434/CHEMRXIV.12518555.V1} {\  (\bibinfo {year}
			{2020}),\ 10.26434/CHEMRXIV.12518555.V1}\BibitemShut {NoStop}%
		\bibitem [{\citenamefont {Ribeiro}\ \emph {et~al.}(2020)\citenamefont
			{Ribeiro}, \citenamefont {Campos-Gonzalez-Angulo}, \citenamefont {Giebink},
			\citenamefont {Xiong},\ and\ \citenamefont {Yuen-Zhou}}]{Ribeiro2020}%
		\BibitemOpen
		\bibfield  {author} {\bibinfo {author} {\bibfnamefont {R.~F.}\ \bibnamefont
				{Ribeiro}}, \bibinfo {author} {\bibfnamefont {J.~A.}\ \bibnamefont
				{Campos-Gonzalez-Angulo}}, \bibinfo {author} {\bibfnamefont {N.~C.}\
				\bibnamefont {Giebink}}, \bibinfo {author} {\bibfnamefont {W.}~\bibnamefont
				{Xiong}},\ and\ \bibinfo {author} {\bibfnamefont {J.}~\bibnamefont
				{Yuen-Zhou}},\ }\bibfield  {title} {\enquote {\bibinfo {title} {{Enhanced
						optical nonlinearities under strong light-matter coupling}},}\ }\href
		{http://arxiv.org/abs/2006.08519} {\  (\bibinfo {year} {2020})},\ \Eprint
		{https://arxiv.org/abs/2006.08519} {arXiv:2006.08519} \BibitemShut {NoStop}%
		\bibitem [{\citenamefont {Li}, \citenamefont {Subotnik},\ and\ \citenamefont
			{Nitzan}(2020)}]{Li2020Water}%
		\BibitemOpen
		\bibfield  {author} {\bibinfo {author} {\bibfnamefont {T.~E.}\ \bibnamefont
				{Li}}, \bibinfo {author} {\bibfnamefont {J.~E.}\ \bibnamefont {Subotnik}},\
			and\ \bibinfo {author} {\bibfnamefont {A.}~\bibnamefont {Nitzan}},\
		}\bibfield  {title} {\enquote {\bibinfo {title} {{Cavity molecular dynamics
						simulations of liquid water under vibrational ultrastrong coupling}},}\
		}\href {https://doi.org/10.1073/pnas.2009272117} {\bibfield  {journal}
			{\bibinfo  {journal} {Proc. Natl. Acad. Sci.}\ }\textbf {\bibinfo {volume}
				{117}},\ \bibinfo {pages} {18324--18331} (\bibinfo {year} {2020})},\ \Eprint
		{https://arxiv.org/abs/2004.04888} {arXiv:2004.04888} \BibitemShut {NoStop}%
		\bibitem [{\citenamefont {Herrera}\ and\ \citenamefont
			{Spano}(2016)}]{Herrera2016}%
		\BibitemOpen
		\bibfield  {author} {\bibinfo {author} {\bibfnamefont {F.}~\bibnamefont
				{Herrera}}\ and\ \bibinfo {author} {\bibfnamefont {F.~C.}\ \bibnamefont
				{Spano}},\ }\bibfield  {title} {\enquote {\bibinfo {title}
				{{Cavity-Controlled Chemistry in Molecular Ensembles}},}\ }\href
		{https://doi.org/10.1103/PhysRevLett.116.238301} {\bibfield  {journal}
			{\bibinfo  {journal} {Phys. Rev. Lett.}\ }\textbf {\bibinfo {volume} {116}},\
			\bibinfo {pages} {238301} (\bibinfo {year} {2016})}\BibitemShut {NoStop}%
		\bibitem [{\citenamefont {Flick}\ \emph {et~al.}(2017)\citenamefont {Flick},
			\citenamefont {Ruggenthaler}, \citenamefont {Appel},\ and\ \citenamefont
			{Rubio}}]{Flick2017}%
		\BibitemOpen
		\bibfield  {author} {\bibinfo {author} {\bibfnamefont {J.}~\bibnamefont
				{Flick}}, \bibinfo {author} {\bibfnamefont {M.}~\bibnamefont {Ruggenthaler}},
			\bibinfo {author} {\bibfnamefont {H.}~\bibnamefont {Appel}},\ and\ \bibinfo
			{author} {\bibfnamefont {A.}~\bibnamefont {Rubio}},\ }\bibfield  {title}
		{\enquote {\bibinfo {title} {{Atoms and Molecules in Cavities, from Weak to
						Strong Coupling in Quantum-Electrodynamics (QED) Chemistry}},}\ }\href
		{https://doi.org/10.1073/pnas.1615509114} {\bibfield  {journal} {\bibinfo
				{journal} {Proc. Natl. Acad. Sci.}\ }\textbf {\bibinfo {volume} {114}},\
			\bibinfo {pages} {3026--3034} (\bibinfo {year} {2017})}\BibitemShut {NoStop}%
		\bibitem [{\citenamefont {Luk}\ \emph {et~al.}(2017)\citenamefont {Luk},
			\citenamefont {Feist}, \citenamefont {Toppari},\ and\ \citenamefont
			{Groenhof}}]{Luk2017}%
		\BibitemOpen
		\bibfield  {author} {\bibinfo {author} {\bibfnamefont {H.~L.}\ \bibnamefont
				{Luk}}, \bibinfo {author} {\bibfnamefont {J.}~\bibnamefont {Feist}}, \bibinfo
			{author} {\bibfnamefont {J.~J.}\ \bibnamefont {Toppari}},\ and\ \bibinfo
			{author} {\bibfnamefont {G.}~\bibnamefont {Groenhof}},\ }\bibfield  {title}
		{\enquote {\bibinfo {title} {{Multiscale Molecular Dynamics Simulations of
						Polaritonic Chemistry}},}\ }\href {https://doi.org/10.1021/acs.jctc.7b00388}
		{\bibfield  {journal} {\bibinfo  {journal} {J. Chem. Theory Comput.}\
			}\textbf {\bibinfo {volume} {13}},\ \bibinfo {pages} {4324--4335} (\bibinfo
			{year} {2017})}\BibitemShut {NoStop}%
		\bibitem [{\citenamefont {Groenhof}\ \emph {et~al.}(2019)\citenamefont
			{Groenhof}, \citenamefont {Climent}, \citenamefont {Feist}, \citenamefont
			{Morozov},\ and\ \citenamefont {Toppari}}]{Groenhof2019}%
		\BibitemOpen
		\bibfield  {author} {\bibinfo {author} {\bibfnamefont {G.}~\bibnamefont
				{Groenhof}}, \bibinfo {author} {\bibfnamefont {C.}~\bibnamefont {Climent}},
			\bibinfo {author} {\bibfnamefont {J.}~\bibnamefont {Feist}}, \bibinfo
			{author} {\bibfnamefont {D.}~\bibnamefont {Morozov}},\ and\ \bibinfo {author}
			{\bibfnamefont {J.~J.}\ \bibnamefont {Toppari}},\ }\bibfield  {title}
		{\enquote {\bibinfo {title} {{Tracking Polariton Relaxation with Multiscale
						Molecular Dynamics Simulations}},}\ }\href
		{https://doi.org/10.1021/acs.jpclett.9b02192} {\bibfield  {journal} {\bibinfo
				{journal} {J. Phys. Chem. Lett.}\ }\textbf {\bibinfo {volume} {10}},\
			\bibinfo {pages} {5476--5483} (\bibinfo {year} {2019})}\BibitemShut {NoStop}%
		\bibitem [{\citenamefont {Sch{\"{a}}fer}\ \emph {et~al.}(2020)\citenamefont
			{Sch{\"{a}}fer}, \citenamefont {Ruggenthaler}, \citenamefont {Rokaj},\ and\
			\citenamefont {Rubio}}]{Schafer2020}%
		\BibitemOpen
		\bibfield  {author} {\bibinfo {author} {\bibfnamefont {C.}~\bibnamefont
				{Sch{\"{a}}fer}}, \bibinfo {author} {\bibfnamefont {M.}~\bibnamefont
				{Ruggenthaler}}, \bibinfo {author} {\bibfnamefont {V.}~\bibnamefont
				{Rokaj}},\ and\ \bibinfo {author} {\bibfnamefont {A.}~\bibnamefont {Rubio}},\
		}\bibfield  {title} {\enquote {\bibinfo {title} {{Relevance of the Quadratic
						Diamagnetic and Self-Polarization Terms in Cavity Quantum
						Electrodynamics}},}\ }\href {https://doi.org/10.1021/acsphotonics.9b01649}
		{\bibfield  {journal} {\bibinfo  {journal} {ACS Photonics}\ }\textbf
			{\bibinfo {volume} {7}},\ \bibinfo {pages} {975--990} (\bibinfo {year}
			{2020})}\BibitemShut {NoStop}%
		\bibitem [{\citenamefont {Mandal}, \citenamefont {{Montillo Vega}},\ and\
			\citenamefont {Huo}(2020)}]{Mandal2020}%
		\BibitemOpen
		\bibfield  {author} {\bibinfo {author} {\bibfnamefont {A.}~\bibnamefont
				{Mandal}}, \bibinfo {author} {\bibfnamefont {S.}~\bibnamefont {{Montillo
						Vega}}},\ and\ \bibinfo {author} {\bibfnamefont {P.}~\bibnamefont {Huo}},\
		}\bibfield  {title} {\enquote {\bibinfo {title} {{Polarized Fock States and
						the Dynamical Casimir Effect in Molecular Cavity Quantum Electrodynamics}},}\
		}\href {https://doi.org/10.1021/acs.jpclett.0c02399} {\bibfield  {journal}
			{\bibinfo  {journal} {J. Phys. Chem. Lett.}\ }\textbf {\bibinfo {volume}
				{11}},\ \bibinfo {pages} {9215--9223} (\bibinfo {year} {2020})}\BibitemShut
		{NoStop}%
		\bibitem [{\citenamefont {Xiang}\ \emph
			{et~al.}(2019{\natexlab{c}})\citenamefont {Xiang}, \citenamefont {Ribeiro},
			\citenamefont {Chen}, \citenamefont {Wang}, \citenamefont {Du}, \citenamefont
			{Yuen-Zhou},\ and\ \citenamefont {Xiong}}]{Xiang2019}%
		\BibitemOpen
		\bibfield  {author} {\bibinfo {author} {\bibfnamefont {B.}~\bibnamefont
				{Xiang}}, \bibinfo {author} {\bibfnamefont {R.~F.}\ \bibnamefont {Ribeiro}},
			\bibinfo {author} {\bibfnamefont {L.}~\bibnamefont {Chen}}, \bibinfo {author}
			{\bibfnamefont {J.}~\bibnamefont {Wang}}, \bibinfo {author} {\bibfnamefont
				{M.}~\bibnamefont {Du}}, \bibinfo {author} {\bibfnamefont {J.}~\bibnamefont
				{Yuen-Zhou}},\ and\ \bibinfo {author} {\bibfnamefont {W.}~\bibnamefont
				{Xiong}},\ }\bibfield  {title} {\enquote {\bibinfo {title} {{State-Selective
						Polariton to Dark State Relaxation Dynamics}},}\ }\href
		{https://doi.org/10.1021/acs.jpca.9b04601} {\bibfield  {journal} {\bibinfo
				{journal} {J. Phys. Chem. A}\ }\textbf {\bibinfo {volume} {123}},\ \bibinfo
			{pages} {5918--5927} (\bibinfo {year} {2019}{\natexlab{c}})}\BibitemShut
		{NoStop}%
		\bibitem [{\citenamefont {Cohen-Tannoudji}, \citenamefont {Dupont-Roc},\ and\
			\citenamefont {Grynberg}(1997)}]{Cohen-Tannoudji1997}%
		\BibitemOpen
		\bibfield  {author} {\bibinfo {author} {\bibfnamefont {C.}~\bibnamefont
				{Cohen-Tannoudji}}, \bibinfo {author} {\bibfnamefont {J.}~\bibnamefont
				{Dupont-Roc}},\ and\ \bibinfo {author} {\bibfnamefont {G.}~\bibnamefont
				{Grynberg}},\ }\href
		{http://www.amazon.de/Photons-Atoms-Introduction-Electrodynamics-Professional/dp/0471184330/ref=sr{\_}1{\_}3?ie=UTF8{\&}qid=1455100150{\&}sr=8-3{\&}keywords=Cohen-Tannoudji+photons}
		{\emph {\bibinfo {title} {{Photons and Atoms: Introduction to Quantum
						Electrodynamics}}}}\ (\bibinfo  {publisher} {Wiley},\ \bibinfo {address} {New
			York},\ \bibinfo {year} {1997})\ pp.\ \bibinfo {pages} {280--295}\BibitemShut
		{NoStop}%
		\bibitem [{\citenamefont {Haugland}\ \emph {et~al.}(2020)\citenamefont
			{Haugland}, \citenamefont {Ronca}, \citenamefont {Kj{\o}nstad}, \citenamefont
			{Rubio},\ and\ \citenamefont {Koch}}]{Haugland2020}%
		\BibitemOpen
		\bibfield  {author} {\bibinfo {author} {\bibfnamefont {T.~S.}\ \bibnamefont
				{Haugland}}, \bibinfo {author} {\bibfnamefont {E.}~\bibnamefont {Ronca}},
			\bibinfo {author} {\bibfnamefont {E.~F.}\ \bibnamefont {Kj{\o}nstad}},
			\bibinfo {author} {\bibfnamefont {A.}~\bibnamefont {Rubio}},\ and\ \bibinfo
			{author} {\bibfnamefont {H.}~\bibnamefont {Koch}},\ }\bibfield  {title}
		{\enquote {\bibinfo {title} {{Coupled Cluster Theory for Molecular
						Polaritons: Changing Ground and Excited States}},}\ }\href
		{https://doi.org/10.1103/PhysRevX.10.041043} {\bibfield  {journal} {\bibinfo
				{journal} {Phys. Rev. X}\ }\textbf {\bibinfo {volume} {10}},\ \bibinfo
			{pages} {041043} (\bibinfo {year} {2020})}\BibitemShut {NoStop}%
		\bibitem [{Note1()}]{Note1}%
		\BibitemOpen
		\bibinfo {note} {As a practical matter, even though the raw value of
			$\protect \cc@accent {"705E}{\setbox \z@ \hbox {\mathsurround \z@ $\textstyle
					q$}\mathaccent "0365{q}}_{k,\lambda }$ is different from the raw value of
			$\protect \cc@accent {"705E}{q}_{k,\lambda }$, the spectrum and the energy of
			the photon can be calculated with either $\protect \cc@accent
			{"705E}{q}_{k,\lambda }$ or $\protect \cc@accent {"705E}{\setbox \z@ \hbox
				{\mathsurround \z@ $\textstyle q$}\mathaccent "0365{q}}_{k,\lambda }$ and the
			same results can be obtained.}\BibitemShut {Stop}%
		\bibitem [{\citenamefont {Feist}, \citenamefont
			{Fern{\'{a}}ndez-Dom{\'{i}}nguez},\ and\ \citenamefont
			{Garc{\'{i}}a-Vidal}(2020)}]{Feist2020}%
		\BibitemOpen
		\bibfield  {author} {\bibinfo {author} {\bibfnamefont {J.}~\bibnamefont
				{Feist}}, \bibinfo {author} {\bibfnamefont {A.~I.}\ \bibnamefont
				{Fern{\'{a}}ndez-Dom{\'{i}}nguez}},\ and\ \bibinfo {author} {\bibfnamefont
				{F.~J.}\ \bibnamefont {Garc{\'{i}}a-Vidal}},\ }\bibfield  {title} {\enquote
			{\bibinfo {title} {{Macroscopic QED for quantum nanophotonics:
						emitter-centered modes as a minimal basis for multiemitter problems}},}\
		}\href {https://doi.org/10.1515/nanoph-2020-0451} {\bibfield  {journal}
			{\bibinfo  {journal} {Nanophoton.}\ }\textbf {\bibinfo {volume} {10}},\
			\bibinfo {pages} {477--489} (\bibinfo {year} {2020})},\ \Eprint
		{https://arxiv.org/abs/2008.02106} {arXiv:2008.02106} \BibitemShut {NoStop}%
		\bibitem [{Note2()}]{Note2}%
		\BibitemOpen
		\bibinfo {note} {Although in principle an external pulse can also directly
			excite cavity photons, this feature is not included in Eq. \protect \textup
			{\hbox {\mathsurround \z@ \protect \normalfont (\ignorespaces \ref
					{eq:EOM_MD_PBC}\unskip \@@italiccorr )}}. Including such an excitation would
			not qualitatively change the simulation results in this manuscript since
			pumping either the molecular or the photonic part equivalently pumps
			polaritons. More importantly, in order to include this feature, one would
			need to introduce a new phenomenological term, namely the effective
			transition dipole of cavity photons, the magnitude of which varies for
			different cavities. Therefore, introducing this term would bring an
			additional manipulatable parameter and would hinder the universality of our
			simulation results: in Sec. \ref {sec:result_PBC} we will show that the
			presented results are universal for cavities with different
			volumes.}\BibitemShut {Stop}%
		\bibitem [{\citenamefont {McQuarrie}(1976)}]{McQuarrie1976}%
		\BibitemOpen
		\bibfield  {author} {\bibinfo {author} {\bibfnamefont {D.~A.}\ \bibnamefont
				{McQuarrie}},\ }\href@noop {} {\emph {\bibinfo {title} {{Statistical
						Mechanics}}}}\ (\bibinfo  {publisher} {Harper-Collins Publish- ers},\
		\bibinfo {address} {New York},\ \bibinfo {year} {1976})\BibitemShut {NoStop}%
		\bibitem [{\citenamefont {Gaigeot}\ and\ \citenamefont
			{Sprik}(2003)}]{Gaigeot2003}%
		\BibitemOpen
		\bibfield  {author} {\bibinfo {author} {\bibfnamefont {M.-P.}\ \bibnamefont
				{Gaigeot}}\ and\ \bibinfo {author} {\bibfnamefont {M.}~\bibnamefont
				{Sprik}},\ }\bibfield  {title} {\enquote {\bibinfo {title} {{Ab Initio
						Molecular Dynamics Computation of the Infrared Spectrum of Aqueous
						Uracil}},}\ }\href {https://doi.org/10.1021/jp034788u} {\bibfield  {journal}
			{\bibinfo  {journal} {J. Phys. Chem. B}\ }\textbf {\bibinfo {volume} {107}},\
			\bibinfo {pages} {10344--10358} (\bibinfo {year} {2003})}\BibitemShut
		{NoStop}%
		\bibitem [{\citenamefont {Habershon}, \citenamefont {Fanourgakis},\ and\
			\citenamefont {Manolopoulos}(2008)}]{Habershon2008}%
		\BibitemOpen
		\bibfield  {author} {\bibinfo {author} {\bibfnamefont {S.}~\bibnamefont
				{Habershon}}, \bibinfo {author} {\bibfnamefont {G.~S.}\ \bibnamefont
				{Fanourgakis}},\ and\ \bibinfo {author} {\bibfnamefont {D.~E.}\ \bibnamefont
				{Manolopoulos}},\ }\bibfield  {title} {\enquote {\bibinfo {title}
				{{Comparison of path integral molecular dynamics methods for the infrared
						absorption spectrum of liquid water}},}\ }\href
		{https://doi.org/10.1063/1.2968555} {\bibfield  {journal} {\bibinfo
				{journal} {J. Chem. Phys.}\ }\textbf {\bibinfo {volume} {129}},\ \bibinfo
			{pages} {074501} (\bibinfo {year} {2008})}\BibitemShut {NoStop}%
		\bibitem [{\citenamefont {Nitzan}(2006)}]{Nitzan2006}%
		\BibitemOpen
		\bibfield  {author} {\bibinfo {author} {\bibfnamefont {A.}~\bibnamefont
				{Nitzan}},\ }\href@noop {} {\emph {\bibinfo {title} {{Chemical Dynamics in
						Condensed Phases: Relaxation, Transfer and Reactions in Condensed Molecular
						Systems}}}}\ (\bibinfo  {publisher} {Oxford University Press},\ \bibinfo
		{address} {New York},\ \bibinfo {year} {2006})\BibitemShut {NoStop}%
		\bibitem [{\citenamefont {Cygan}, \citenamefont {Romanov},\ and\ \citenamefont
			{Myshakin}(2012)}]{Cygan2012}%
		\BibitemOpen
		\bibfield  {author} {\bibinfo {author} {\bibfnamefont {R.~T.}\ \bibnamefont
				{Cygan}}, \bibinfo {author} {\bibfnamefont {V.~N.}\ \bibnamefont {Romanov}},\
			and\ \bibinfo {author} {\bibfnamefont {E.~M.}\ \bibnamefont {Myshakin}},\
		}\bibfield  {title} {\enquote {\bibinfo {title} {{Molecular Simulation of
						Carbon Dioxide Capture by Montmorillonite Using an Accurate and Flexible
						Force Field}},}\ }\href {https://doi.org/10.1021/jp3007574} {\bibfield
			{journal} {\bibinfo  {journal} {J. Phys. Chem. C}\ }\textbf {\bibinfo
				{volume} {116}},\ \bibinfo {pages} {13079--13091} (\bibinfo {year}
			{2012})}\BibitemShut {NoStop}%
		\bibitem [{\citenamefont {Darwent}(1970)}]{darwent1970bond}%
		\BibitemOpen
		\bibfield  {author} {\bibinfo {author} {\bibfnamefont {B.~d.}\ \bibnamefont
				{Darwent}},\ }\href@noop {} {\emph {\bibinfo {title} {Bond dissociation
					energies in simple molecules}}}\ (\bibinfo  {publisher} {U.S. National Bureau
			of Standards},\ \bibinfo {year} {1970})\BibitemShut {NoStop}%
		\bibitem [{\citenamefont {Mart{\'{i}}nez}\ \emph {et~al.}(2009)\citenamefont
			{Mart{\'{i}}nez}, \citenamefont {Andrade}, \citenamefont {Birgin},\ and\
			\citenamefont {Mart{\'{i}}nez}}]{Martinez2009}%
		\BibitemOpen
		\bibfield  {author} {\bibinfo {author} {\bibfnamefont {L.}~\bibnamefont
				{Mart{\'{i}}nez}}, \bibinfo {author} {\bibfnamefont {R.}~\bibnamefont
				{Andrade}}, \bibinfo {author} {\bibfnamefont {E.~G.}\ \bibnamefont
				{Birgin}},\ and\ \bibinfo {author} {\bibfnamefont {J.~M.}\ \bibnamefont
				{Mart{\'{i}}nez}},\ }\bibfield  {title} {\enquote {\bibinfo {title}
				{{PACKMOL: A package for building initial configurations for molecular
						dynamics simulations}},}\ }\href {https://doi.org/10.1002/jcc.21224}
		{\bibfield  {journal} {\bibinfo  {journal} {J. Comput. Chem.}\ }\textbf
			{\bibinfo {volume} {30}},\ \bibinfo {pages} {2157--2164} (\bibinfo {year}
			{2009})}\BibitemShut {NoStop}%
		\bibitem [{\citenamefont {Kapil}\ \emph {et~al.}(2019)\citenamefont {Kapil},
			\citenamefont {Rossi}, \citenamefont {Marsalek}, \citenamefont {Petraglia},
			\citenamefont {Litman}, \citenamefont {Spura}, \citenamefont {Cheng},
			\citenamefont {Cuzzocrea}, \citenamefont {Mei{\ss}ner}, \citenamefont
			{Wilkins}, \citenamefont {Helfrecht}, \citenamefont {Juda}, \citenamefont
			{Bienvenue}, \citenamefont {Fang}, \citenamefont {Kessler}, \citenamefont
			{Poltavsky}, \citenamefont {Vandenbrande}, \citenamefont {Wieme},
			\citenamefont {Corminboeuf}, \citenamefont {K{\"{u}}hne}, \citenamefont
			{Manolopoulos}, \citenamefont {Markland}, \citenamefont {Richardson},
			\citenamefont {Tkatchenko}, \citenamefont {Tribello}, \citenamefont {{Van
					Speybroeck}},\ and\ \citenamefont {Ceriotti}}]{Kapil2019}%
		\BibitemOpen
		\bibfield  {author} {\bibinfo {author} {\bibfnamefont {V.}~\bibnamefont
				{Kapil}}, \bibinfo {author} {\bibfnamefont {M.}~\bibnamefont {Rossi}},
			\bibinfo {author} {\bibfnamefont {O.}~\bibnamefont {Marsalek}}, \bibinfo
			{author} {\bibfnamefont {R.}~\bibnamefont {Petraglia}}, \bibinfo {author}
			{\bibfnamefont {Y.}~\bibnamefont {Litman}}, \bibinfo {author} {\bibfnamefont
				{T.}~\bibnamefont {Spura}}, \bibinfo {author} {\bibfnamefont
				{B.}~\bibnamefont {Cheng}}, \bibinfo {author} {\bibfnamefont
				{A.}~\bibnamefont {Cuzzocrea}}, \bibinfo {author} {\bibfnamefont {R.~H.}\
				\bibnamefont {Mei{\ss}ner}}, \bibinfo {author} {\bibfnamefont {D.~M.}\
				\bibnamefont {Wilkins}}, \bibinfo {author} {\bibfnamefont {B.~A.}\
				\bibnamefont {Helfrecht}}, \bibinfo {author} {\bibfnamefont {P.}~\bibnamefont
				{Juda}}, \bibinfo {author} {\bibfnamefont {S.~P.}\ \bibnamefont {Bienvenue}},
			\bibinfo {author} {\bibfnamefont {W.}~\bibnamefont {Fang}}, \bibinfo {author}
			{\bibfnamefont {J.}~\bibnamefont {Kessler}}, \bibinfo {author} {\bibfnamefont
				{I.}~\bibnamefont {Poltavsky}}, \bibinfo {author} {\bibfnamefont
				{S.}~\bibnamefont {Vandenbrande}}, \bibinfo {author} {\bibfnamefont
				{J.}~\bibnamefont {Wieme}}, \bibinfo {author} {\bibfnamefont
				{C.}~\bibnamefont {Corminboeuf}}, \bibinfo {author} {\bibfnamefont {T.~D.}\
				\bibnamefont {K{\"{u}}hne}}, \bibinfo {author} {\bibfnamefont {D.~E.}\
				\bibnamefont {Manolopoulos}}, \bibinfo {author} {\bibfnamefont {T.~E.}\
				\bibnamefont {Markland}}, \bibinfo {author} {\bibfnamefont {J.~O.}\
				\bibnamefont {Richardson}}, \bibinfo {author} {\bibfnamefont
				{A.}~\bibnamefont {Tkatchenko}}, \bibinfo {author} {\bibfnamefont {G.~A.}\
				\bibnamefont {Tribello}}, \bibinfo {author} {\bibfnamefont {V.}~\bibnamefont
				{{Van Speybroeck}}},\ and\ \bibinfo {author} {\bibfnamefont {M.}~\bibnamefont
				{Ceriotti}},\ }\bibfield  {title} {\enquote {\bibinfo {title} {{i-PI 2.0: A
						universal force engine for advanced molecular simulations}},}\ }\href
		{https://doi.org/10.1016/j.cpc.2018.09.020} {\bibfield  {journal} {\bibinfo
				{journal} {Comput. Phys. Commun.}\ }\textbf {\bibinfo {volume} {236}},\
			\bibinfo {pages} {214--223} (\bibinfo {year} {2019})}\BibitemShut {NoStop}%
		\bibitem [{\citenamefont {Plimpton}(1995)}]{Plimpton1995}%
		\BibitemOpen
		\bibfield  {author} {\bibinfo {author} {\bibfnamefont {S.}~\bibnamefont
				{Plimpton}},\ }\bibfield  {title} {\enquote {\bibinfo {title} {{Fast Parallel
						Algorithms for Short-Range Molecular Dynamics}},}\ }\href
		{https://doi.org/10.1006/jcph.1995.1039} {\bibfield  {journal} {\bibinfo
				{journal} {J. Comput. Phys.}\ }\textbf {\bibinfo {volume} {117}},\ \bibinfo
			{pages} {1--19} (\bibinfo {year} {1995})}\BibitemShut {NoStop}%
		\bibitem [{\citenamefont {Li}(2020)}]{TELi2020Github}%
		\BibitemOpen
		\bibfield  {author} {\bibinfo {author} {\bibfnamefont {T.~E.}\ \bibnamefont
				{Li}},\ }\href {https://github.com/TaoELi/cavity-md-ipi} {\enquote {\bibinfo
				{title} {{Cavity Molecular Dynamics Simulations Tool Sets}},}\ }\bibinfo
		{howpublished} {https://github.com/TaoELi/cavity-md-ipi} (\bibinfo {year}
		{2020})\BibitemShut {NoStop}%
		\bibitem [{\citenamefont {Seki}, \citenamefont {Grunwaldt},\ and\ \citenamefont
			{Baiker}(2009)}]{Seki2009}%
		\BibitemOpen
		\bibfield  {author} {\bibinfo {author} {\bibfnamefont {T.}~\bibnamefont
				{Seki}}, \bibinfo {author} {\bibfnamefont {J.-D.}\ \bibnamefont
				{Grunwaldt}},\ and\ \bibinfo {author} {\bibfnamefont {A.}~\bibnamefont
				{Baiker}},\ }\bibfield  {title} {\enquote {\bibinfo {title} {{In Situ
						Attenuated Total Reflection Infrared Spectroscopy of Imidazolium-Based
						Room-Temperature Ionic Liquids under “Supercritical” CO 2}},}\ }\href
		{https://doi.org/10.1021/jp800424d} {\bibfield  {journal} {\bibinfo
				{journal} {J. Phys. Chem. B}\ }\textbf {\bibinfo {volume} {113}},\ \bibinfo
			{pages} {114--122} (\bibinfo {year} {2009})}\BibitemShut {NoStop}%
		\bibitem [{Note3()}]{Note3}%
		\BibitemOpen
		\bibinfo {note} {Note that consistent polaritonic relaxation dynamics can
			also be captured by evaluating the square norm of the molecular total dipole
			moment ($\left \langle |\protect \boldsymbol {\mu }_S(t)|^2\right \rangle -
			|\left \langle \protect \boldsymbol {\mu }_S(t)\right \rangle
			|^2$).}\BibitemShut {Stop}%
		\bibitem [{\citenamefont {Xiong}(2020)}]{WXiong2020private}%
		\BibitemOpen
		\bibfield  {author} {\bibinfo {author} {\bibfnamefont {W.}~\bibnamefont
				{Xiong}},\ }\href@noop {} {}\bibinfo {howpublished} {private communication}
		(\bibinfo {year} {2020})\BibitemShut {NoStop}%
		\bibitem [{\citenamefont {Wang}\ \emph {et~al.}(2020)\citenamefont {Wang},
			\citenamefont {Seidel}, \citenamefont {Nagarajan}, \citenamefont {Chervy},
			\citenamefont {Genet},\ and\ \citenamefont {Ebbesen}}]{Wang2020}%
		\BibitemOpen
		\bibfield  {author} {\bibinfo {author} {\bibfnamefont {K.}~\bibnamefont
				{Wang}}, \bibinfo {author} {\bibfnamefont {M.}~\bibnamefont {Seidel}},
			\bibinfo {author} {\bibfnamefont {K.}~\bibnamefont {Nagarajan}}, \bibinfo
			{author} {\bibfnamefont {T.}~\bibnamefont {Chervy}}, \bibinfo {author}
			{\bibfnamefont {C.}~\bibnamefont {Genet}},\ and\ \bibinfo {author}
			{\bibfnamefont {T.~W.}\ \bibnamefont {Ebbesen}},\ }\bibfield  {title}
		{\enquote {\bibinfo {title} {{Large optical nonlinearity enhancement under
						electronic strong coupling}},}\ }\href {http://arxiv.org/abs/2005.13325} {\
			(\bibinfo {year} {2020})},\ \Eprint {https://arxiv.org/abs/2005.13325}
		{arXiv:2005.13325} \BibitemShut {NoStop}%
		\bibitem [{\citenamefont {Fu}\ \emph {et~al.}(2012)\citenamefont {Fu},
			\citenamefont {Wang}, \citenamefont {Long}, \citenamefont {Yang},
			\citenamefont {Lu}, \citenamefont {Hetsch}, \citenamefont {Susha},\ and\
			\citenamefont {Rogach}}]{Fu2012}%
		\BibitemOpen
		\bibfield  {author} {\bibinfo {author} {\bibfnamefont {M.}~\bibnamefont
				{Fu}}, \bibinfo {author} {\bibfnamefont {K.}~\bibnamefont {Wang}}, \bibinfo
			{author} {\bibfnamefont {H.}~\bibnamefont {Long}}, \bibinfo {author}
			{\bibfnamefont {G.}~\bibnamefont {Yang}}, \bibinfo {author} {\bibfnamefont
				{P.}~\bibnamefont {Lu}}, \bibinfo {author} {\bibfnamefont {F.}~\bibnamefont
				{Hetsch}}, \bibinfo {author} {\bibfnamefont {A.~S.}\ \bibnamefont {Susha}},\
			and\ \bibinfo {author} {\bibfnamefont {A.~L.}\ \bibnamefont {Rogach}},\
		}\bibfield  {title} {\enquote {\bibinfo {title} {{Resonantly enhanced optical
						nonlinearity in hybrid semiconductor quantum dot – metal nanoparticle
						structures}},}\ }\href {https://doi.org/10.1063/1.3683548} {\bibfield
			{journal} {\bibinfo  {journal} {Appl. Phys. Lett.}\ }\textbf {\bibinfo
				{volume} {100}},\ \bibinfo {pages} {063117} (\bibinfo {year}
			{2012})}\BibitemShut {NoStop}%
		\bibitem [{\citenamefont {Rivera}\ \emph {et~al.}(2017)\citenamefont {Rivera},
			\citenamefont {Rosolen}, \citenamefont {Joannopoulos}, \citenamefont
			{Kaminer},\ and\ \citenamefont {Solja{\v{c}}i{\'{c}}}}]{Rivera2017}%
		\BibitemOpen
		\bibfield  {author} {\bibinfo {author} {\bibfnamefont {N.}~\bibnamefont
				{Rivera}}, \bibinfo {author} {\bibfnamefont {G.}~\bibnamefont {Rosolen}},
			\bibinfo {author} {\bibfnamefont {J.~D.}\ \bibnamefont {Joannopoulos}},
			\bibinfo {author} {\bibfnamefont {I.}~\bibnamefont {Kaminer}},\ and\ \bibinfo
			{author} {\bibfnamefont {M.}~\bibnamefont {Solja{\v{c}}i{\'{c}}}},\
		}\bibfield  {title} {\enquote {\bibinfo {title} {{Making two-photon processes
						dominate one-photon processes using mid-IR phonon polaritons}},}\ }\href
		{https://doi.org/10.1073/pnas.1713538114} {\bibfield  {journal} {\bibinfo
				{journal} {Proc. Natl. Acad. Sci.}\ }\textbf {\bibinfo {volume} {114}},\
			\bibinfo {pages} {13607--13612} (\bibinfo {year} {2017})}\BibitemShut
		{NoStop}%
		\bibitem [{\citenamefont {Scandolo}\ and\ \citenamefont
			{Bassani}(1992)}]{Scandolo1992}%
		\BibitemOpen
		\bibfield  {author} {\bibinfo {author} {\bibfnamefont {S.}~\bibnamefont
				{Scandolo}}\ and\ \bibinfo {author} {\bibfnamefont {F.}~\bibnamefont
				{Bassani}},\ }\bibfield  {title} {\enquote {\bibinfo {title} {{Nonlinear sum
						rules: The three-level and the anharmonic-oscillator models}},}\ }\href
		{https://doi.org/10.1103/PhysRevB.45.13257} {\bibfield  {journal} {\bibinfo
				{journal} {Phys. Rev. B}\ }\textbf {\bibinfo {volume} {45}},\ \bibinfo
			{pages} {13257--13261} (\bibinfo {year} {1992})}\BibitemShut {NoStop}%
		\bibitem [{Note4()}]{Note4}%
		\BibitemOpen
		\bibinfo {note} {Note that this weak enhancement of nonlinearity might be
			responsible for the relative short LP lifetime as found in Fig. \ref
			{fig:lifetime_all}a.}\BibitemShut {Stop}%
		\bibitem [{\citenamefont {Gross}\ and\ \citenamefont
			{Haroche}(1982)}]{Gross1982}%
		\BibitemOpen
		\bibfield  {author} {\bibinfo {author} {\bibfnamefont {M.}~\bibnamefont
				{Gross}}\ and\ \bibinfo {author} {\bibfnamefont {S.}~\bibnamefont
				{Haroche}},\ }\bibfield  {title} {\enquote {\bibinfo {title} {{Superradiance:
						An Essay on the Theory of Collective Spontaneous Emission}},}\ }\href
		{https://doi.org/10.1016/0370-1573(82)90102-8} {\bibfield  {journal}
			{\bibinfo  {journal} {Phys. Rep.}\ }\textbf {\bibinfo {volume} {93}},\
			\bibinfo {pages} {301--396} (\bibinfo {year} {1982})}\BibitemShut {NoStop}%
		\bibitem [{\citenamefont {Li}\ \emph {et~al.}(2018)\citenamefont {Li},
			\citenamefont {Nitzan}, \citenamefont {Sukharev}, \citenamefont {Martinez},
			\citenamefont {Chen},\ and\ \citenamefont {Subotnik}}]{Li2018Spontaneous}%
		\BibitemOpen
		\bibfield  {author} {\bibinfo {author} {\bibfnamefont {T.~E.}\ \bibnamefont
				{Li}}, \bibinfo {author} {\bibfnamefont {A.}~\bibnamefont {Nitzan}}, \bibinfo
			{author} {\bibfnamefont {M.}~\bibnamefont {Sukharev}}, \bibinfo {author}
			{\bibfnamefont {T.}~\bibnamefont {Martinez}}, \bibinfo {author}
			{\bibfnamefont {H.-T.}\ \bibnamefont {Chen}},\ and\ \bibinfo {author}
			{\bibfnamefont {J.~E.}\ \bibnamefont {Subotnik}},\ }\bibfield  {title}
		{\enquote {\bibinfo {title} {{Mixed Quantum--Classical Electrodynamics:
						Understanding Spontaneous Decay and Zero--Point Energy}},}\ }\href
		{https://doi.org/10.1103/PhysRevA.97.032105} {\bibfield  {journal} {\bibinfo
				{journal} {Phys. Rev. A}\ }\textbf {\bibinfo {volume} {97}},\ \bibinfo
			{pages} {032105} (\bibinfo {year} {2018})}\BibitemShut {NoStop}%
		\bibitem [{\citenamefont {Goggin}\ and\ \citenamefont
			{Milonni}(1988)}]{Goggin1988}%
		\BibitemOpen
		\bibfield  {author} {\bibinfo {author} {\bibfnamefont {M.~E.}\ \bibnamefont
				{Goggin}}\ and\ \bibinfo {author} {\bibfnamefont {P.~W.}\ \bibnamefont
				{Milonni}},\ }\bibfield  {title} {\enquote {\bibinfo {title} {{Driven Morse
						oscillator: Classical chaos, quantum theory, and photodissociation}},}\
		}\href {https://doi.org/10.1103/PhysRevA.37.796} {\bibfield  {journal}
			{\bibinfo  {journal} {Phys. Rev. A}\ }\textbf {\bibinfo {volume} {37}},\
			\bibinfo {pages} {796--806} (\bibinfo {year} {1988})}\BibitemShut {NoStop}%
	\end{thebibliography}
	%

\end{document}